\documentclass[11pt]{article}
\pdfoutput=1
\usepackage{jcappub,bm,natbib,float,comment,placeins}
\usepackage[utf8]{inputenc}
\DeclareUnicodeCharacter{2212}{-}
\def\be{\begin{equation}}
\def\ee{\end{equation}}
\def\ba#1\ea{\begin{align}#1\end{align}}
\newcommand{\vs}{\nonumber\\}

\renewcommand{\v}[1]{\bm{#1}}
\newcommand{\vx}{\v{x}}

\newcommand{\vk}{\v{k}}
\newcommand{\vp}{\v{p}}
\newcommand{\vq}{\v{q}}

\newcommand{\refeq}[1]{Eq.~(\ref{eq:#1})}          
\newcommand{\refeqs}[2]{Eqs.~(\ref{eq:#1})--(\ref{eq:#2})}          
\newcommand{\reffig}[1]{figure~\ref{fig:#1}} 
\newcommand{\reffigs}[2]{figures~(\ref{fig:#1})--(\ref{fig:#2})}
\newcommand{\refFig}[1]{Figure~\ref{fig:#1}}
          
\newcommand{\refsec}[1]{section~\ref{sec:#1}}   
 \newcommand{\refsecs}[2]{sections~\ref{sec:#1}--\ref{sec:#2}}       
\newcommand{\refSec}[1]{Section~\ref{sec:#1}}          
\newcommand{\refapp}[1]{Appendix~\ref{app:#1}}

\newcommand{\Om}{\Omega_m}
\newcommand{\Ob}{\Omega_b}

\renewcommand{\d}{\delta}

\def\iMpch{\,h\,{\rm Mpc}^{-1}}

\newcommand{\dm}{\delta_m}
\newcommand{\dpk}{\delta_{\rm pk}}

\def\O{\mathcal{O}}
\def\Mpch{\,h^{-1}{\rm Mpc}}

\newcommand{\blapl}{b_{\lapl\d}}
\newcommand{\cseff}{C^2_{s,{\rm eff}}}

\def\N{\mathcal{N}}
\def\lapl{\nabla^2}

\newcommand{\e}{\operatorname{e}}

\newcommand{\<}{\left\langle}
\renewcommand{\>}{\right\rangle}

\def\knl{k_\text{NL}}
\def\Plin{P_\text{L}}

\title{A robust measurement of the first higher-derivative bias of dark matter halos}

\author[a,b,c,d,e]{Titouan Lazeyras,}
\author[b]{Fabian~Schmidt}
\affiliation[a]{SISSA, Via Bonomea 265, 34136 Trieste, Italy}
\affiliation[b]{Max-Planck-Institut f\"ur Astrophysik, Karl-Schwarzschild-Str. 1, 85748 Garching, Germany}
\affiliation[c]{Kavli Institute for Theoretical Physics, University of California at Santa Barbara, 1102 Kohn Hall, Santa Barbara, CA 93106, United States}
\affiliation[d]{INFN, Sezione di Trieste, Via Bonomea 265, 34136 Trieste, Italy}
\affiliation[e]{IFPU, Institute for Fundamental Physics of the Universe, via Beirut 2, 34151, Trieste, Italy}
\emailAdd{tlazeyra@sissa.it, fabians@mpa-garching.mpg.de}
\abstract{We present a new simulation technique in which any chosen mode $\vk$ of the density contrast field can be amplified by an amplitude $\Delta$. These \textit{amplified-mode simulations} allow us to study the response of the halo density field to a long-wavelength mode other than the DC mode. In this sense they are a generalization of the separate-universe simulations to finite-wavelength modes. In particular, we use these simulations to obtain robust measurements of the first higher-derivative bias of dark matter halos $\blapl$. We find a negative bias at all mass considered, roughly following the $-R_L^2(M)$ relation, the Lagrangian radius of halos squared, as naively expected. We compare our results with those obtained from a fit to the 1-loop halo-matter power spectrum, as well as with the recent results from Abidi and Baldauf (2018), and to the prediction from the peak theory. }

\begin{document}

\maketitle
\flushbottom

%%%%%%%%%%%%%%%%%%%%%%%%%%%%%%%%%%%%%%%%%%%%%%%%%%%%%%%%%%%%%%%%%%%%%%%%%%%%%
%%%%%%%%%%%%%%%%%%%%%%%%%%%%%%%%%%%%%%%%%%%%%%%%%%%%%%%%%%%%%%%%%%%%%%%%%%%%%
\section{Introduction}
\label{sec:intro}

The large-scale distribution of dark matter halos is one of the key
ingredients of the theoretical description of large-scale structure (LSS).  
Since most observed tracers of LSS, such as galaxies, reside in halos,
their statistics is determined by those of halos on large scales.  
In the context of perturbation theory, the statistics of halos are written in terms of bias parameters $b_O$ multiplying operators $O$ constructed out of the matter density field $\dm$ and the tidal field $K_{ij}$ (see \cite{biasreview} for a recent review)
\be
\d_h(\vx,\tau)=\sum_{O} b_O(\tau) O(\vx,\tau),
\label{eq:biasexpO}
\ee 
where $\d_h$ is the fractional number density perturbation of a given halo sample.

Operators entering \refeq{biasexpO} can be divided into two categories: those which include exactly two net derivatives of the gravitational potential field $\Phi$, and \textit{higher-derivative} operators which include four or more net derivatives of $\Phi$. Physically, these higher-derivative operators encapsulate the fact that halo formation involves the collapse of matter from a finite region in space, and thus, \refeq{biasexpO} cannot be completely local on all scales. Starting from a simple linear relation 
\be
\d_h(\vx,\tau)\supset b_1 \dm(\vx,\tau)\,,
\ee
the way to incorporate the deviation from perfect locality of halo formation is to replace the local operators $\dm(\vx,\tau)$ with a functional \cite{Matsubara:1999, Coles:2007}
\be
b_1(\tau)\dm(\vx,\tau) \rightarrow \int d^3\v{y} F(\v{y},\tau)\dm(\vx+\v{y},\tau) ,
\ee
where $F(\v{y},\tau)$ is a kernel that is in general time dependent but has to be independent of $\vx$ by homogeneity of the Universe. Performing a formal series expansion of $\dm$ around $\vx$ leads to
\be
b_1(\tau)\dm(\vx,\tau) \rightarrow b_1(\tau)\dm(\vx,\tau)+\blapl(\tau)\lapl\dm(\vx,\tau)+\cdots \, ,
\label{eq:bloctoblapl}
\ee
where statistical isotropy demands the absence of any preferred directions with which the derivative operators could be contracted. Hence the leading higher-derivative operator involves the Laplacian of $\dm(\vx,\tau)$, and the associated bias parameter has dimension $[{\rm length}]^2$. Its magnitude is expected to be of the order of $R_*^2$, where $R_*$ is the scale of the spatial support of the kernel $F(\v{y},\tau)$, which we identify with the nonlocality scale of the tracer. For halos of mass $M$, this is given by their Lagrangian radius $R_L(M)$. In Fourier space, the term proportional to $\blapl$ corresponds to a ``scale-dependent bias'' $−\blapl k^2 \d$. However, let us emphasize that this is an expansion in powers of $k^2$ , rather than a general function $f(k)$, which is how the term ``scale-dependent bias'' is sometimes interpreted. To make this distinction clear, we will use the term higher-derivative bias throughout.

The peak model introduced by \cite{Bardeen:1985} allows for a theoretical prediction of $\blapl$. Indeed, in this model, the real-space peak-matter 2-point correlation function can be derived in closed form in Lagrangian space, yielding analytic forms for the local Lagrangian bias $b^L_1$ and the peak bias $b_{01}^L$. We can then use a model for velocity bias (which arises from the correlation between linear velocities and density gradients, and reflects the fact that large-scale flows are more likely to be directed towards peaks than to be oriented randomly) to compute the Eulerian peak bias $b^E_{01}$ and,  by taking into account the effect of the smoothing filter we can get an analytic prediction for the Eulerian higher-derivative bias.

On the measurements side, the first constraints on $\blapl$ have been placed by studies testing the scale dependence of bias on large scales \cite{Scherrer:1997,Mann:1997}. More recent measurements include those of \cite{Fujita:2016} who found a value of $3[R_L(M)]^2$ (but only ruled out $\blapl=0$ at the $\sim 1.3\sigma$ level). On the other hand, \cite{Angulo:2015} quote values for $\blapl$ that are much smaller than $[R_L(M)]^2$. Very recently, \cite{Abidi:2018} measured $\blapl$ by fitting the 1-loop halo-matter power spectrum with one free parameter. They found a result consistent within errors with $-[R_L(M)]^2/\alpha$ with $\alpha \approx 2-3$. Thus, there is still large uncertainty on the magnitude of Eulerian higher-derivative bias for halos. This can be measured more easily in Lagrangian space, using either the halo-matter power spectrum \cite{Elia:2012, Baldauf:2014} or the projection method of \cite{Paranjape:2013,Biagetti:2013}. In particular, \cite{Elia:2012, Baldauf:2014} measured the so-called peak bias $b_{01}^L$, which contributes to $-b^L_{\lapl\d}$ along with the leading contribution from the filtering kernel, and obtained $b_{01}^L \approx 2[R_L(M)]^2$ for halos with mass $M \geq 8 \times 10^{12} h^{−1} M_\odot$ , with only a weak departure from the simple $[R_L(M)]^2$ scaling. However, as we explained above, it is not possible to relate Lagrangian higher-derivative biases to their Eulerian counterpart without using a model for the velocity bias.

In this paper, we propose a new technique to measure directly $\blapl$ in Eulerian space using so-called \textit{amplified-mode simulations}. The idea is to enhance a single mode $\vk_0$ by adding a modulation $\Delta\cos(\vk_0\cdot\vx)$ in the initial conditions for the density field of a gravitation-only N-body simulation, which, assuming linear growth, translates to $B_m\cos(\vk_0\cdot\vx)$ at low redshift. This is a generalization of the separate universe simulations introduced in, e.g. \cite{Wagner:2014}, to non-DC modes with finite wavenumber. This enhancement amplifies the contribution of $\blapl\lapl\dm$ in the bias expansion allowing for a clear detection of the linear higher-derivative bias parameter. More precisely, if the mode $\vk_0$ is chosen to be small enough that linear theory still applies today, the same wavelength mode should be observed in the halo density field with a different amplitude, i.e. $\d_h$ receives a contribution of the type $B_h\cos(\vk_0\cdot\vx)$. Since the linear relation between $\dm$ and $\d_h$ is given Fourier space by
\be
\d_h(\vk)=[b_1-k^2\blapl]\dm(\vk) ,
\label{eq:linbias}
\ee
one can measure $\blapl$ from a suite of amplified-mode simulations by measuring the ratio $B_h/\Delta$ for diverse values of $\vk_0$ and fit a second order polynomial to this relation.  

This paper is organized as follows: in \refsec{amsims} we describe in more details the idea of amplified-mode simulations (\refsec{idea}), how to estimate $\blapl$ from them (\refsec{estbk}) and the higher-order corrections one needs to consider (\refsecs{corr}{hcorr}). We present our set of simulations and shortly explain the halo finding procedure in \refsec{sims}. \refSec{predictions} describes how to obtain the same results from the 1-loop power spectrum in perturbation theory (\refsec{powerspec}) and reviews some aspects of the peak theory and how the higher-derivative bias can be computed using this model (\refsec{peaks}). Finally, \refsec{results} presents and discusses our results, and we conclude in \refsec{concl}. The appendices present some checks of our implementation of the simulations (\refapp{checksims}), detailed computation of various quantities in amplified-mode simulations (\refapp{deltaandP}), some considerations about measuring the effective sound speed for matter $\cseff$ (\refapp{cseff}) and comparison between our amplified-mode technique and 1-loop power spectrum results (\refapp{Phm1loop}). Throughout this paper we adopt the same cosmology as in \cite{Lazeyras:2015}, i.e. a flat $\Lambda {\rm CDM}$ cosmology with $\Om = 0.27$, $h = 0.7$, $\Ob h^2=0.023$ and $\mathcal{A}_s = 2.2 \cdot 10^{-9}$.

%%%%%%%%%%%%%%%%%%%%%%%%%%%%%%%%%%%%%%%%%%%%%%%%%%%%%%%%%%%%%%%%%%%%%%%%%%%%%
%%%%%%%%%%%%%%%%%%%%%%%%%%%%%%%%%%%%%%%%%%%%%%%%%%%%%%%%%%%%%%%%%%%%%%%%%%%%%
\section{Amplified-mode simulations}
\label{sec:amsims}

In this section, we introduce in more details the idea behind amplified-mode simulations. Here and in the following we drop the redshift argument from the equations since the results can be applied at any $z$.

\subsection{Theoretical idea}
\label{sec:idea}

The principle of amplified-mode simulations is to superimpose a plane wave of linear amplitude $\Delta$ and wavenumber $\vk_0$ to the initial random density field $\d^{(1)}(\vx)\big|_{\Delta=0}$ coming from sampling the power spectrum in the absence of enhancement (we use the notation $\delta^{(1)}$ to denote the linear density field). Since the density field has to be real in configuration space it is given by
\be
\d^{(1)}(\vx)=\d^{(1)}(\vx)\big|_{\Delta=0}+\Delta {\rm Re}(\e^{i\vk_0\cdot\vx})=\d^{(1)}(\vx)\big|_{\Delta=0}+\Delta\cos(\vk_0\cdot\vx).
\label{eq:enhancedd}
\ee
Here, we have set the phase of the plane wave to zero. The value of the phase is arbitrary, given homogeneity and isotropy of the background. 
As mentioned above, the effect of the amplified mode is to amplify the contribution of $\blapl \lapl \d_m$ in the bias expansion, allowing for a clear detection of the higher-derivative bias factor. Since we are only interested in measuring the linear higher-derivative bias parameter, we choose the mode $\vk_0$ to be on sufficiently large scales that linear theory still applies at redshift zero.

The implementation of the amplified mode in N-body simulations is straightforward since one only needs to modify the initial distribution of particles to incorporate the plane wave before running the simulation in a traditional way. Hence no modification needs to be done to the integration scheme. Using the fact that the density field is discrete and periodic in both configuration and Fourier space, the Fourier transform of \refeq{enhancedd} is
\be
\d^{(1)}(\vk)=\d^{(1)}(\v{n}k_F)=\d^{(1)}(\v{n}k_F)\big|_{\Delta=0}+\frac{\Delta L^3}{2}(\d^K_{\v{n},\v{m}}+\d^K_{\v{-n},\v{m}}),
\label{eq:enhanceddk}
\ee
where $k_F=(2\pi)/L$ is the modulus of the fundamental mode of the simulation box ($L$ is the one dimensional comoving box size), $\v{n}$ a vector of integers, $\v{m}$ another vector of integers such that $\vk_0=\v{m}k_F$ and $\d^K_{\v{n},\v{m}}$ the Kronecker delta.\footnote{We see explicitly  in this last expression that the reality condition on the configuration space density field involves the hermitianity of the Fourier space field with each of the two modes $\v{m}$ and $-\v{m}$ being amplified.} Thus, after sampling the density field from the power spectrum in Fourier space, one simply needs to add a factor of $(\Delta L^3)/2$ at the desired wavenumber and ensure hermitianity of the Fourier space density field. The positions and velocities of particles are then obtained from the 2LPT displacement field and the resulting distribution of particles given as an input to the cosmological simulation code. The integration scheme is then carried out using a standard N-body code without any further modifications to obtain the late-time particle distribution. In \refapp{checksims} we present some detailed tests to verify our implementation.

\subsection{Estimating the higher-derivative bias}
\label{sec:estbk} 

We present here an estimator for $B_h$ the amplified mode amplitude in the halo density field as well as the procedure to obtain $\blapl$ from this estimator. We also discuss the higher-order corrections that we neglect in \refsec{hcorr}. 
In this section, we absorb the small random contribution from sampling the power spectrum at $k_0$, $\Plin(k_0)\big|_{\Delta=0}$, in $\Delta$. Hence the Lagrangian real space density field is now given by
\be
\d^{(1)}(\vx) = \Delta\,\cos(k_0 x)\,.
\ee
The halo density field can be written in terms of the halo density $n_h$ as
\be
\d_h(\vx) = \frac{n_h(\vx)}{\bar n_h} - 1 = B_h\,\cos(k_0 x)\,,\quad B_h = b(k_0) B_m = [b_1 - b_{\nabla^2\d} k_0^2 + \cdots ] B_m\,,
\ee
where, $B_m$ is the amplitude as measured in the late-time matter field and $\bar{n}_h$ is the mean halo density.
We consider halos within a fixed mass range, but drop the mass argument since the results can be applied to any halo selection. 
Thus, we want to estimate $B_h$.  Given the Gaussian nature of $\d_m$ and $\d_h$ at linear order, we can simply use a $\chi^2$ given by
\be
\chi^2 = \sum_{x_i} \frac1{\N^2} \left[n_h(\vx_i) - \bar n_h\left(1 + B_h \cos(k_0 x_i) \right) \right]^2\,,
\ee
where $\vx_i$ is the position of the $i^{th}$ halo, $x_i$ the $x$ component of $\vx_i$ and $\N$ is a noise term which we assume to be constant in space (since we assume that the modulation of $n_h$ on the scale $k_0$ to be small).  The least-squares estimator for $B_h$ is then
\ba
\frac{\partial}{\partial B_h} \chi^2(B_h) \stackrel{!}{=} 0 \quad\Leftrightarrow\quad
\sum_{x_i} \cos(k_0 x_i) \left[n_h(\vx_i) - \bar n_h\left(1 + B_h \cos(k_0 x_i) \right) \right] = 0\,,
\ea
which yields
\be
\hat B_h = \frac{\sum_{x_i} \cos(k_0 x_i) [n_h(\vx_i) - \bar n_h]}{\bar n_h \sum_{x_i} \cos^2(k_0 x_i)}\,.
\ee
This can be implemented by simply summing over the halo positions $\vx_i$, weighted by factors of $\cos(k_0 x_i)$.
Further, if we assume that halos are approximately uniformly distributed (see the linear approximation above), the cosine-average over a constant vanishes, while the denominator yields
\be
\sum_{x_i} \cos^2(k_0 x_i) \to \int_0^L \frac{dx}{L} \cos^2(k_0 x) = \frac12 \, .
\ee
We then obtain
\be
\hat B_h = \frac{2 \sum_{\rm halos} \cos(k_0 x_i)}{N_{\rm halos}}\,,
\label{eq:hatB}
\ee
where $N_{\rm halos}$ is the total number of halos.  

The same estimator can be applied to matter to obtain $\hat B_m$. One simply needs to replace $B_h \rightarrow B_m$, $B_m \rightarrow \Delta$ in the above derivation, and replace the halo density $n_h$ and total number of halos $N_{\rm halos}$ by the matter density $\rho_m$ and total number of particles $N$. 

To get $\blapl$ in practice we use the Fourier space relation \refeq{linbias} applied to our estimator. In order to partially cancel cosmic variance we actually compute the mean between the positive and negative amplitude results for each realization, i.e.
\begin{align} 
\frac{\hat B^i_h(k_0,+\Delta,M)-\hat B^i_h(k_0,-\Delta,M)}{2\Delta} & = b_1(M)-(\blapl(M)+b_1(M)C^2_{\rm s, eff}) k_0^2 \nonumber \\
&\equiv b_1(M)-b_{\lapl\d}^{\rm eff}(M)k_0^2 ,
\label{eq:blininpractice}
\end{align}
where $i$ denotes the $i^{\rm th}$ realization, $C^2_{\rm s, eff}$ is the scaled effective sound speed of the dark matter fluid, and we use the results of \cite{Lazeyras:2015} for the linear bias $b_1(M)$\footnote{It is crucial for the ratio in \refeq{blininpractice} to be computed for results from the same realization (i.e. the same seed for the random generation of the initial particle distribution) in order to cancel the effects of the random phase at $\vk_0$.}. We used the superscript ``eff'' to emphasize that we took the ratio with respect to the linear amplitude $\Delta$ and we neglected the $4^{\rm th}$ order term. We then simply fit a second-order polynomial in $k_0$ to this ratio to get an estimate for $\blapl^{\rm eff}(M)$. 

Finally, \refeq{blininpractice} involves the effective sound speed for matter $C^2_{\rm s, eff}$ that we will measure from the standard perturbation theory (SPT) 1-loop matter power spectrum prediction ($P_{\rm 1-loop}$) as
\be
\frac{P_{mm}(k)-P_{\rm 1-loop}(k)}{\Plin(k)}=-2C^2_{s,{\rm eff}}k^2 \, ,
\label{eq:cPSPT}
\ee
where $P_{mm}$ is the matter power spectrum as measured in simulations and $\Plin$ the linear one. We can then subtract the product $b_1 \cseff$ from $\blapl^{\rm eff}$ to obtain $\blapl(M)$.

%%%%%%%%%%%%%%%%%%%%%%%%%%%%%%%%%%%%%%%%%%%%%%%%%%%%%%%%%%%%%%%%%%%%%%%%%%%%%
\subsection{Coupling between short and long wavelengths modes}
\label{sec:corr}

In addition to what has been discussed in the previous section, there are nontrivial couplings between long and short wavelength modes that cannot be ignored. We present here the 1-loop calculation needed to compute them. We wish to stay concise and presents the details in \refapp{1loopdelta}.

We want to compute the nonlinear matter and halo fields in the amplified-mode case. Throughout, we assume the infinite-volume limit. Then, the linear density field in the amplified-mode simulations in Fourier space is modified to
\be
\d^{(1)}(\vk) = \d_s^{(1)}(\vk) + \frac12 \Delta (2\pi)^3\left[ e^{i\phi} \d_D(\vk-\vk_0) + e^{-i\phi}\d_D(\vk+\vk_0)\right]\,,
\label{eq:d1mod}
\ee
where $\phi$ is the phase of the amplified-mode and we introduced the notation $\d_s=\d|_{\Delta=0}$ for shortness. 
The two Dirac delta functions ensure that the matter density field remains real, which requires $\d^{(1)}(-\vk) = \d^{(1)*}(\vk)$. In the following, we will set $\phi=0$ without loss of generality. We now consider the halo density field. In perturbation theory, the nonlinear halo density field at the mode $\vk_0$ can be written as
\ba
\d_h(\vk_0) =\:& \left[b_1 - b^{\rm eff}_{\lapl\d} k_0^2 \right] \d^{(1)}(\vk_0) \vs
& + \sum_{n=2}^\infty \int_{\vp_1} \cdots \int_{\vp_n} (2\pi)^3 \d_D(\vk_0 - \vp_{1\cdots n})F_n^{(h)}(\vp_1,\cdots,\vp_n) \d^{(1)}(\vp_1)\cdots \d^{(1)}(\vp_n)\,,
\label{eq:dPT}
\ea
where $\int_{\vp} \equiv \int d^3\vp/(2\pi)^3$ and $F_n^{(h)}$ are the fully symmetrized kernels of the halo density field, which we will describe below. We now insert \refeq{d1mod} into \refeq{dPT} and evaluate the result up to cubic order. The estimator applied to the halo density field is defined as the symmetric difference
\be
\widehat{\frac{{\rm d} \d_h}{{\rm d}\Delta}} \equiv \frac{1}{2\Delta} \left[
\d_h(\vk_0)\Big|_\Delta - \d_h(\vk_0)\Big|_{-\Delta}
\right]\,.
\label{eq:estimator}
\ee
As we show in \refapp{1loopdelta}, the final result for this quantity at cubic order and when averaging over many small modes (so that we can replace the small-scale modes with their ensemble average) is 
\ba
\widehat{\frac{{\rm d} \d_h}{{\rm d}\Delta}} =\:& \left[b_1 - (b_{\lapl\d} + b_1 C_{s,\rm eff}^2) k_0^2 + \O(k_0^4) \right] \frac12 \, (2\pi)^3\d_D(\v{0}) \vs
& + \frac32 \int_{\vp} \Plin(p) F_3^{(h)}(\vp,-\vp,\vk_0) (2\pi)^3\d_D(\v{0})\vs
& + \frac38 (2\pi)^3 \d_D(\v{0})\Delta^2 F_3^{(h)}(\vk_0,-\vk_0,\vk_0) \,.
\label{eq:dh3dPS}
\ea
Now everything is multiplied by the same factor $(2\pi)^3 \d_D(\v{0})$ (which simply gives $L_{\rm box}^3$ when restoring box normalization). We are interested in the terms in the first line. However, we see that there are further contributions from the second and third lines. The cubic kernel in the configuration $F_3^{(h)}(\vp,-\vp,\vk_0)$ is precisely what appears in the 1-loop halo power spectrum (e.g., \cite{assassi/etal}). There are two contributions: first, the cubic order matter kernel $b_1 F_3$ multiplied by the linear bias. Second, there is a contribution from quadratic and cubic bias terms. We have
\ba
F_3^{(h)}(\vp,-\vp,\vk_0) =\:& b_1 F_3(\vp,-\vp,\vk_0) + \frac43 \left( b_{K^2} + \frac25 b_{\rm td}\right) \left[\frac{[\vp\cdot(\vk-\vp)]^2}{p^2 |\vk-\vp|^2} - 1\right] F_2(\vk,-\vp)\,.
\label{eq:Fh3}
\ea
In the limit $p \ll k_0$, $F_3^{(h)}(\vp,-\vp,\pm\vk_0)$ scales as $(k_0/p)^2$. The integrand in the second line of \refeq{dh3d} peaks around $p\sim \knl$, and hence we expect this contribution to be of order $(k_0/\knl)^2 \Delta$, which is not negligible compared to $b_{\lapl\d} k_0^2 \Delta$ except possibly for the most massive halos. 
This kernel involves the bias combination $b_{K^2}+(2/5)b_{\rm td}$, which was recently measured by \cite{Lazeyras:2017} using an optimal estimator for the trispectrum (see their figure 2 which shows $(5/2)b_{K^2}+b_{\rm td}$). We hence use their result multiplied by $2/5$ in \refeq{Fh3}. On the other hand, the last line in \refeq{dh3dPS} is multiplied by an extra factor $\Delta^2$ which allows us to neglect it.

Hence the final expression allowing us to measure $\blapl$ becomes
\ba
2 \cdot \widehat{\frac{{\rm d} \d_h}{{\rm d}\Delta}}(\vk_0) -  3 \int_{\vp} \Plin(p) F_3^{(h)}(\vp,-\vp,\vk_0) =\:& b_1 - (b_{\lapl\d} + b_1 C_{s,\rm eff}^2) k_0^2  \, , \vs
\equiv\: & b_1 -\blapl^{\rm eff} k_0^2.
\label{eq:blaplfinal}
\ea

%%%%%%%%%%%%%%%%%%%%%%%%%%%%%%%%%%%%%%%%%%%%%%%%%%%%%%%%%%%%%%%%%%%%%%%%%%%%%
\subsection{Higher order corrections}
\label{sec:hcorr}

In this section we discuss the various higher-order corrections that we do not take into account and we show that they can indeed be neglected. 

First of all there is the obvious  $\Delta^2$ contribution in the last line of \refeq{dh3dPS}. Since we choose for $\Delta$ a linearly extrapolated value of 0.05 at $z=0$ it is clear that this term can be neglected. As long as corrections of order $\Delta^2$ are negligible, the results for $\blapl$ do not depend on the precise choice of $\Delta$.

Next, there are the 2-loop contributions that we neglected in our calculation at 1-loop order (terms up to cubic order in the linear density field) leading to \refeq{blaplfinal}. We note that showing these contributions to be negligible is also of importance for the estimator of $\blapl$ using the 1-loop halo-matter power spectrum that we will present in \refsec{powerspec}. There are several terms to consider and we will discuss them one after the other. They are shown in figure 1 of \cite{Baldauf:2015aha} and we will follow the same nomenclature as that reference. The first one is the $(3-3)^\text{I}$ contribution (i.e. two tadpoles). This is simply of the order of the $(1-3)$ contribution squared, i.e. 
\be
(3-3)^\text{I} \sim \left( \int_{\vp} F^{(h)}_3(\vp,-\vp,\vk) \Plin(p)  \right)^2 \sim \left(\frac{k}{k_{\rm NL}}\right)^4 \, ,
\label{eq:33}
\ee
where the last approximation is valid in the soft limit $\vk \rightarrow 0$ (where $k_{\rm NL}$ is the nonlinear scale at a given redshift). In the same limit the (1-3) contribution scales as $(k/k_{\rm NL})^2$, and the $(3-3)^\text{I}$ one can hence be neglected. Then there are the $(3-3)^\text{II}$ and $(2-4)$ terms. These in fact simply sum up together to the integral of the 1-loop power spectrum
\be
(3-3)^\text{II}+(2-4) \sim \int_{\vp} F^{(h)}_3(\vp,-\vp,\vk) P_{1-{\rm loop}}(p) \sim \left(\frac{k}{k_{\rm NL}}\right)^2 \, ,
\label{eq:tbd}
\ee
in the soft limit again. The scaling of these terms is hence the same as the one of the 1-loop terms we consider. However, this contribution has to be renormalized with counterterms (in the same way as $\cseff$ appears at 1-loop), and will not contribute to the 1-loop result in the end.
Finally, there is the $(1-5)$ contribution
\be
(1-5) \sim \int_{\vp} \int_{\vp '} F^{(h)}_5(\vp,-\vp,\vp ', -\vp ', \vk) \Plin(p) \Plin(p')  \sim \left(\frac{k}{k_{\rm NL}}\right)^2 \, ,
\label{eq:15}
\ee
in the double soft limit \cite{abolhasani/mirbabayi/pajer:2016}. The scaling of this term is again the same as the one of the 1-loop terms. However, as in the case of the $(3-3)^\text{II}$ and $(2-4)$ terms, this contribution has to be renormalized with counterterms and will not contribute to the 1-loop result in the end. It can hence also be discarded.
We conclude that all 2-loop corrections scale as $(k_0^4/k_{\rm NL})^4$, compared to the $k_0^2$ scaling of the desired higher-derivative contribution, provided that $k_0$ is sufficiently smaller than $k_{\rm NL} \sim 0.25 \iMpch$ at $z=0$.

There is a further constraint on $k_0$ from the amplitude of the higher-derivative biases $\propto \partial^4 \d$ that we neglect. Estimating the latter to have a coefficient of order $R_L^4$ (recall that we expect $\blapl \propto R_L^2$), we obtain 
\ba
\frac{(k_0 R_L)^4}{(k_0 R_L)^2}=\:& (k_0 R_L)^2
\stackrel{!}{\ll} 1\,.
\label{eq:req2}
\ea
Thus, by choosing a value of $k_0 \ll 1/R_L$ and $k_0 \ll k_{\rm NL}$, and making $\Delta$ sufficiently small to neglect the terms proportional to $\Delta^2$, the contribution $\propto b_{\lapl\d}$ can be made to be the leading contribution. 

Finally we note that technically, our measurement using the estimator $\hat B_h$ in \refeq{hatB} corresponds to a version of the ``scatter-plot'' technique to measure bias. That is, we calculate the weighted number of halos, corresponding to a plane-wave filter, for a range of values of the corresponding weighted matter density. As described in detail in Sec.~4.2 of \cite{biasreview}, this technique exactly recovers the bias parameters relating the halo-matter moment (defined with the same filter) to the matter moments. Apart from the choice of filtering kernel, which is usually a spherical or cubic tophat filter but chosen to be plane-wave here, there is one further difference in our application of the technique: the value of the matter density is not random, but chosen deterministically as $\Delta$. This means that we cancel cosmic variance to leading order.

%%%%%%%%%%%%%%%%%%%%%%%%%%%%%%%%%%%%%%%%%%%%%%%%%%%%%%%%%%%%%%%%%%%%%%%%%%%%%
%%%%%%%%%%%%%%%%%%%%%%%%%%%%%%%%%%%%%%%%%%%%%%%%%%%%%%%%%%%%%%%%%%%%%%%%%%%%%
\section{Simulations and halo finding}
\label{sec:sims}

We present here the details of our set of simulations. We also provide a quick outline of the halo finding procedure. 

We arbitrarily align the plane wave in the $x$ direction and choose $\Delta=0.05$ at redshift zero for the linear amplitude. We then run simulations where we amplify the modes $k_0=\{k_F,2k_F,3k_F,4k_F,5k_F, 8k_F, 10k_F\}$, and amplitude $\pm\Delta$ for each $k_0$ value. We choose a comoving box size $L=500 \Mpch$ and number of particles $N=512^3$. These last two parameters are the same as for the ``highres'' set of simulations of \cite{Lazeyras:2015} who computed the local bias parameters from separate universe simulations, and yield a mass resolution $m_p=7\cdot10^{10}h^{-1}M_\odot$. Finally we ran 48 realizations of each simulation and initialized them with 2LPT at $z=49$. We refer to this set of simulations as amplified-mode simulations. In particular we refer to the fiducial set corresponding to no amplification as L500.

Furthermore, in order to cross-check our results with constraints from the power spectrum, we use another set of two simulations without amplified-mode, and with the same cosmological parameters but box size $L=2400\Mpch$ and $N=1536^3$ particles. This allows us to increase the signal-to-noise ratio on large-scales. The mass resolution in this set is $m_p=2.9 \cdot 10^{11}h^{-1}M_\odot$ and we refer to it as L2400.

The halo finding procedure is same as the one used in \cite{Lazeyras:2015}. Halos are identified at $z=0$, 0.5 and 1 using the spherical overdensity halo finder Amiga Halo Finder (AHF) \cite{Gill:2004, Knollmann:2009} with an overdensity threshold $200\rho_m$ for the halo definition ($\rho_m$ is the background density). We bin the mass range of halos in 11 tophat bins of width 0.2 in logarithmic scale centered from $\log M =12.55$ to $\log M=14.55$, where $\log$ is the base 10 logarithm.  Hence the lowest  mass bin is centered on halos with around 51 particles, with a lower limit around 40 particles. We refer the reader to \cite{Lazeyras:2015} for more details and the justification of our choices.  

Before moving on we shortly come back to the condition given by \refeq{req2}. For our simulation parameters, $10k_F=0.126 \, h \, {\rm Mpc}^{-1}$ and $1/R_L$ is between 0.089 and 0.483 $h \, {\rm Mpc}^{-1}$ so that \refeq{req2} is satisfied for $k \leq 8 k_F (5 k_F)$ for objects of in bins $\log M= 14.35 (14.55)$ respectively and up to $10 k_F$ for all less massive objects. This defines the range of $k_0$ values that we will use for the fit as a function of halo mass.

%%%%%%%%%%%%%%%%%%%%%%%%%%%%%%%%%%%%%%%%%%%%%%%%%%%%%%%%%%%%%%%%%%%%%%%%%%%%%
\section{Other measurements and predictions}
\label{sec:predictions}

In this section we present how the same results can be obtained from the 1-loop power spectrum in SPT, as well as predictions from the peak model for $\blapl$.

\subsection{Power spectrum measurements}
\label{sec:powerspec}

We start by describing how to measure the higher-derivative bias parameter from the 1-loop halo-matter power spectrum. This will provide a good cross-check of our results. This has already been done in \cite{Fujita:2016, Abidi:2018}.

The one-loop halo-matter power spectrum is given by (see e.g. \cite{biasreview} and references therein)
\ba
P_{hm}^{1-{\rm loop}}(k) & = b_1\left[P_{mm}^{1-{\rm loop}}(k)-2C_{s,{\rm eff}}^2 k^2 \Plin(k)\right] \vs
& + b_2 \int_{\vp}F_2(\vk-\vp,\vp)\Plin(p)\Plin(|\vk-\vp |) \vs
& + 2b_{K^2}\int_{\vp}F_2(\vk-\vp,\vp)\left[\left(\frac{\vk-\vp}{|\vk-\vp |}\cdot\frac{\vp}{p}\right)^2-\frac13\right]\Plin(p)\Plin(|\vk-\vp |) \vs
& + 4\left(b_{K^2}+\frac25 b_{\rm td}\right)\Plin(k)\int_{\vp}F_2(\vk,-\vp)\left[\frac{[\vp \cdot (\vk-\vp)]^2}{p^2|\vk-\vp |^2}-1\right]\Plin(p) \vs
& - \blapl k^2 \Plin(k)\,,
\label{eq:Phm1loop}
\ea
where we have neglected the stochastic contribution which is proportional to $k^2$ and expected to be smaller than the 1-loop order terms. Ref. \cite{Lazeyras:2017} recently measured the bias parameters $b_1$, $b_2$, $b_{K^2}$, and $b_{\rm td}$ for the same cosmology as in this work, which allows us to fit the halo-matter power spectrum measured from simulations with a single free parameter to obtain a measurement of $\blapl$.

In practice we use the L500 set of simulations to fit \refeq{Phm1loop} up to $k_{\rm max}=0.15 \iMpch$ at all redshift. We choose this value for the same reason as for $\cseff$ (see \refapp{cseff}) as well as to have a maximum $k$ roughly matching those of our amplified-mode simulations. We then follow the same procedure as outlined in \cite{Lazeyras:2017} to obtain robust errorbars. Mainly, we first use a bootstrap technique to obtain errorbars on each data point as a function of the wavenumber $k$. We then use these errorbars to weight the points when fitting, and bootstrap the fit in order to obtain errorbars on the final result for $\blapl$. We then use the two simulations of the L2400 set to obtain the final mean value of $\blapl$ using errorbars scaled by the total effective volume (both on the data at each $k$ and on $\blapl$ itself). This means that, knowing the $1\sigma$ error from the L500 set ([$\sigma(\blapl)]_{\rm L500}$), we infer the one for the L2400 set as
\be
[\sigma(\blapl)]_{\rm L2400}=\sqrt{\frac{V_{\rm L500}}{V_{\rm L2400}}}\,[\sigma(\blapl)]_{\rm L500}\, ,
\label{eq:errorscaling}
\ee
where $V_{\rm L500}=48\cdot500^3 (\Mpch)^3$ and $V_{\rm L500}=2\cdot2400^3 (\Mpch)^3$. For more details and justification about this procedure, we refer the reader to section 3.1 of \cite{Lazeyras:2017}.

\subsection{Prediction from peak theory}
\label{sec:peaks} 

In this section, we introduce how the higher-derivative bias can be estimated from the peak model first introduced in \citep{Bardeen:1985}. Since the peak theory has already been extensively discussed in the literature, we refrain from giving a detailed description of this model here (we refer the reader to the original paper \citep{Bardeen:1985}).
Notice that the apparition of a scale-dependent bias for peaks, as well as the concept of velocity bias (that we will introduce shortly) were first pointed out in  \cite{Desjacques:2008jj, Desjacques:2010:D81} and further studied in, e.g. \cite{Desjacques:2010, Baldauf:2014, Baldauf:2016}. We define the following spectral moments for a generic window function $W$
\be
\sigma^2_i=\int\frac{d^3\vk}{(2\pi)^3}k^{2i}\Plin(k)W^2,
\label{eq:sigmas}
\ee
where $\Plin(k)$ is again the linear power spectrum, as well as the spectral shape parameter
\be
\gamma=\frac{\sigma^2_1}{\sigma_0\sigma_2}.
\label{eq:gamma}
\ee

In the peak model, halos are in one-to-one correspondence with peaks of the Lagrangian density field. This assumption is expected to hold for halos with masses above a few $M_*$, where $M_*$ is the typical mass of halos that collapsed at redshift $z$. In Fourier space, the density of peaks $\dpk$ is written in terms of the density field filtered on some scale $R$ as
\be
\dpk(\vk)=(b^E_{10}+b^E_{01}k^2)\d_R(\vk),
\label{eq:dpeak}
\ee
where $b^E_{10}=1+b^L_{10}=b^E_1$ is the local halo bias, $b^E_{01}$ contributes to $\blapl$ and $\d_R(\vk)=\d(\vk)W(\vk)$. We choose for the filter the effective window function introduced in \cite{Chan:2015}
\be 
W(kR)=W_G(kR/5)W_{TH}(kR)=\e^{-(kR/5)^2/2}\frac{3}{(kR)^3}[\sin(kR)-kR\cos(kR)],
\label{eq:effwindow}
\ee
with $W_G$ and $W_{TH}$ the gaussian and tophat filters respectively. Expanding  $W$ in a Taylor series we get
\be
\dpk(\vk)=b_{10}^E\d(\vk)-\left(\frac{3}{25}R^2b_{10}^E-b^E_{01}\right)k^2\d(\vk)+\mathcal{O}(k^4).
\label{eq:dpeak2}
\ee

To get an expression for $b_{10}^E$ and $b_{01}^E$ we start from their Lagrangian counterpart given by
\begin{align}
b^L_{10} &=  \frac{1}{\sigma_0}\left(\frac{\nu_c-\gamma J}{1-\gamma^2}\right),\label{eq:b10l} \\ 
b_{01}^L &= \frac{1}{\sigma_2}\left(\frac{J-\gamma\nu_c}{1-\gamma^2}\right),
\label{eq:b01l}
\end{align}
where $\nu_c=\frac{\d_c}{\sigma_0}$ with $\d_c=1.686$ the critical threshold for collapse, $J=G_1(\gamma,\gamma\nu_c)/G_0(\gamma,\gamma\nu_c)$ is the mean peak curvature and 
\be
G_n(\gamma,\nu)=\int_0^\infty dx x^nf(x)\frac{\e^{-(x-\nu)^2/2(1-\gamma^2)}}{\sqrt{2\pi(1-\gamma^2)}},
\label{eq:G}
\ee
with $f(x)$ the function defined in Eq. (A.15) of \citep{Bardeen:1985}. To obtain the Eulerian counterpart of \refeqs{b10l}{b01l}, we must take into account the so-called velocity bias, i.e. the fact that large-scale flows are more likely to be orientated towards peaks than in random directions. The velocity bias is defined in terms of the peak and linear matter displacement fields as \cite{Desjacques:2008jj, Desjacques:2010:D81}
\be
\v{s}_{\rm pk}(\vk)=\left(1-\frac{\sigma_0^2}{\sigma_1^2}k^2\right)W(\vk)\v{s}_m(\vk)\equiv c_{v, {\rm pk}}(k)\v{s}_m(\vk), 
\label{eq:velbias}
\ee
where the displacement fields map the Lagrangian ($\vq$) to Eulerian ($\vx$) positions of matter particles and peaks of the density field
\be
\vx_i=\vq_i+\v{s}_i,
\ee
with the subscript $i$ standing for peaks (pk) and matter (m), respectively. Integrating the continuity equation $\partial \dpk(\vk)/\partial \tau=-\nabla \cdot \v{v}_{\rm pk}(\vk)$ and evaluating the result in Fourier space then yields
\begin{align}
b_{10}^E &=1+D(z)b_{10}^L , \label{eq:b10E} \\
b_{01}^E &= -\frac{\sigma_0^2}{\sigma_1^2}+D(z)b_{01}^L ,
\label{eq:b01E}
\end{align}
where $D(z)$ is the linear growth factor at redshift $z$ normalized so that $D(z_0)=1$ for halos collapsing at redshift $z_0$. In these last two expressions it is important to understand that halo collapse is meant to happen at redshift $z_0$. Hence, to compute the Eulerian biases one should fix $z=z_0$ and compute the spectral moments \refeq{sigmas} at this redshift (including the ones entering the Lagrangian biases), keeping $D=1$. 

We can hence plug \refeqs{b10E}{b01E} in \refeq{dpeak2} and obtain the higher-derivative bias parameter in the peak model
\be
\blapl^{\rm pk}=\left[\frac{3}{25}b_{10}^ER^2-b_{01}^E\right],
\label{eq:blaplpeak}
\ee 
with the dependence on the mass and the redshift being implicit.

%%%%%%%%%%%%%%%%%%%%%%%%%%%%%%%%%%%%%%%%%%%%%%%%%%%%%%%%%%%%%%%%%%%%%%%%%%%%%
\section{Results and discussion}
\label{sec:results}

We now turn to presenting our results for the effective sound speed of matter, as well as for the leading higher-order derivative bias. 

\subsection{Effective matter sound speed}
\label{sec:matterbias}

We show here results for the effective sound speed of matter using the 1-loop power spectrum for matter \refeq{cPSPT}. To ensure that the 1-loop power spectrum accurately describes the matter power spectrum we perform the fit to $k_{\rm max} \ll k_{\rm NL}$, where $k_{\rm NL}$ is the nonlinear scale defined such that 
\be
\frac{k_{\rm NL}^3 P_{mm}(k_{\rm NL})}{2\pi^2}=1. 
\label{eq:knl}
\ee
For our cosmology, $k_{\rm NL}=0.22, \, 0.32$ and $0.47 \, h/{\rm Mpc}$ at redshift $0,\, 0.5$ and $1$ respectively. Hence we perform fits up to $k_{\rm max}=0.15 \, h/{\rm Mpc}$ at all redshift (see \refapp{cseff} for a more detailed justification of this choice).

\begin{figure}
\centering
\includegraphics[scale=0.5]{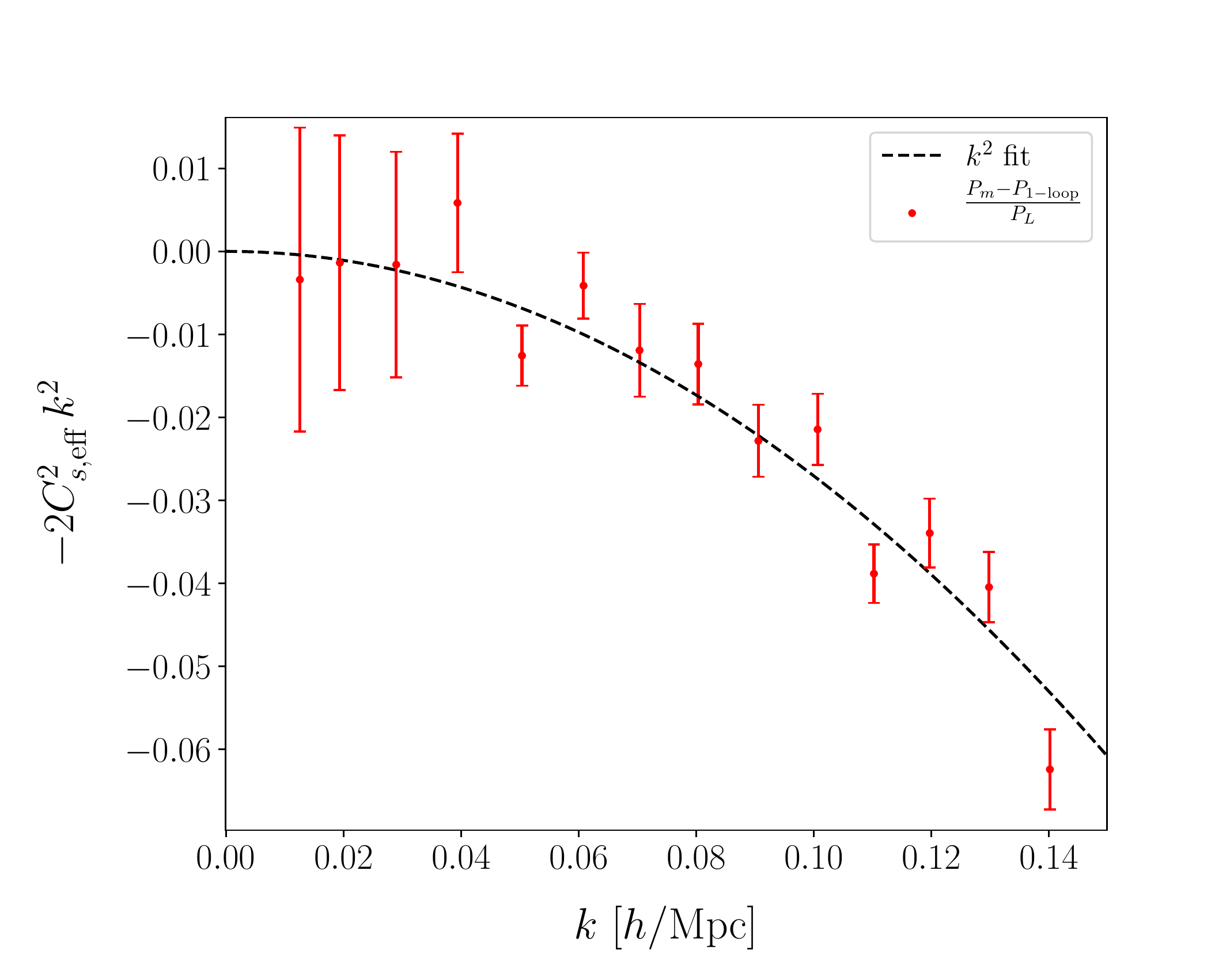}
\caption{The ratio $(P_{mm}-P_{\rm 1-loop})/\Plin$ as a function of the wavenumber $k$ used to determine the effective sound speed for matter $C_{s, {\rm eff}}^2$ at $z=0$. The dashed line shows the best fit when we fit up to $k_{\rm max}=0.15 \iMpch$.} 
\label{fig:cofk}
\end{figure}

The relation corresponding to \refeq{cPSPT} at $z=0.0$ as a function of $k$ is presented in \reffig{cofk}. The red points were obtained from the L2400 set of simulations where the 1-sigma errorbars have been rescaled from the ones obtained with the L500 set, as explained in \refsec{powerspec}. Fitting a second order polynomial to this relation we determine $C^2_{\rm s,eff}=1.31\pm 0.06 \, (\Mpch)^2$ at $z=0$. We also get  $C^2_{\rm s,eff}=0.65\pm 0.05 \, (\Mpch)^2$ and $C^2_{\rm s,eff}=0.33\pm 0.04 \, (\Mpch)^2$ at  $z=0.5$ and $1$ respectively. 

Several works presented results for $\cseff$ measured in a similar way, including \cite{Carrasco:2013sva} who obtained $\cseff=1.6 \; ({\rm Mpc}/h)^2$,  and \cite{Angulo:2015} who quote a value of $C^2_{s,{\rm eff}}=2.31 \pm 0.02 \; ({\rm Mpc}/h)^2$. Ref. \cite{Baldauf:2015aha} also obtained results at various redshifts quoting, e.g. a value of $0.98 \; ({\rm Mpc}/h)^2$ at $z=0$. They furthermore provided a very detailed discussion about the effect of the 2-loop contributions that we neglect for this measurement. We elaborate on these results and show further tests about this quantity in \refapp{cseff}.

\subsection{Halo higher-derivative bias}
\label{sec:halobias} 

\begin{figure}
\centering
\includegraphics[scale=0.5]{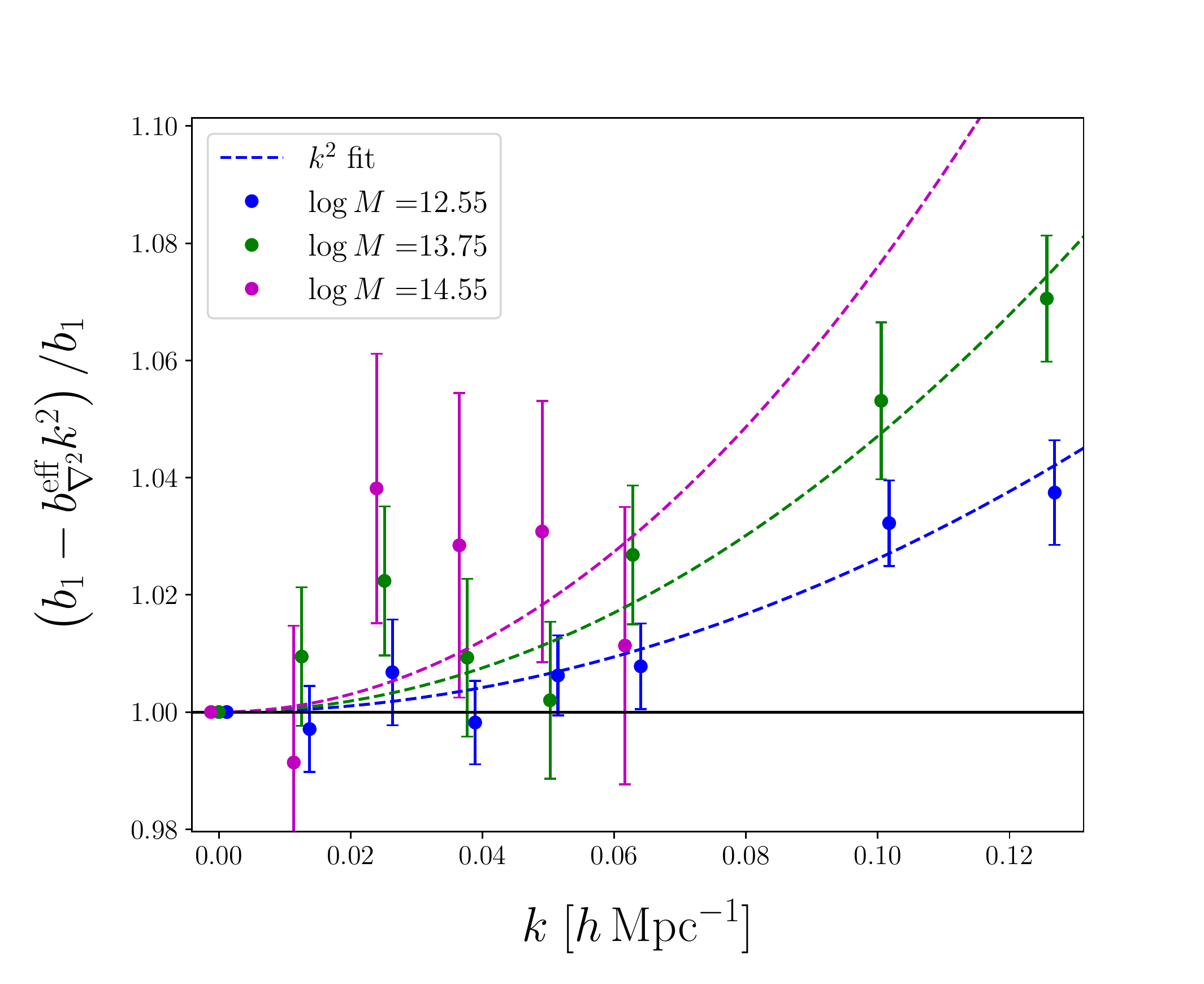}\
\caption{The ratio $\left(b_1-b_{\nabla^2}^{\mathrm{eff}}k^2\right)/b_1$ as a function of $k$ for three mass bins as indicated by the color coding at redshift 0.  We have already corrected for the coupling between short and long wavelength modes so that this ratio corresponds to \refeq{blaplfinal} divided by $b_1$. The dotted lines present the best fit used to determine $\blapl$. We have divided by $b_1$ to avoid the well known mass dependence that would result in a shift between the curves along the $y$-axis. The blue and magenta points have been slightly displaced horizontally for clarity.}
\label{fig:bofk}
\end{figure}

We now turn to results for $b_{\nabla^2\d}(M)$. We first present the mean ratio in \refeq{blaplfinal} (divided by $b_1$ to remove the mass dependent shift along the vertical axis) as a function of $k$ for three mass bins (color coded) in \reffig{bofk}. The dashed lines on this figure show the best $k^2$ fits that are then used to determine $\blapl$ together with the results from the previous section for $C^2_{\rm s, eff}$ as
\be
\blapl(M)=b_{\lapl\d}^{\rm eff}(M)-b_1(M)C^2_{\rm s, eff}.
\label{eq:bfromblin}
\ee
In order to respect the condition in \refeq{req2} we use $k_{\rm max}=5 \, (8) k_F$ as a maximum wavenumber for the fit at $\log M = 14.35$ (14.55) respectively. For all lower mass bins we use the full $k$-range up to 10$k_F$. We checked the dependence of our results for $\blapl$ on the maximum $k$ used for the fit, and found results consistent within $1\sigma$ errorbars, with of course a higher signal-to-noise ratio when including higher wavenumbers in the fit. This indicates that the 2-loop effects discussed in \refsec{hcorr} are sufficiently small to not significantly bias the measurement of $\blapl$. In \refapp{cseff}, we confirm that $C^2_{\rm s,eff}$, which similarly receives 2-loop corrections, also does not show any significant scale dependence on the scales used for constraining $\blapl$, $k < 10k_F=0.126 \, h \, {\rm Mpc}^{-1}$.

\begin{figure}
\centering
\includegraphics[scale=0.47]{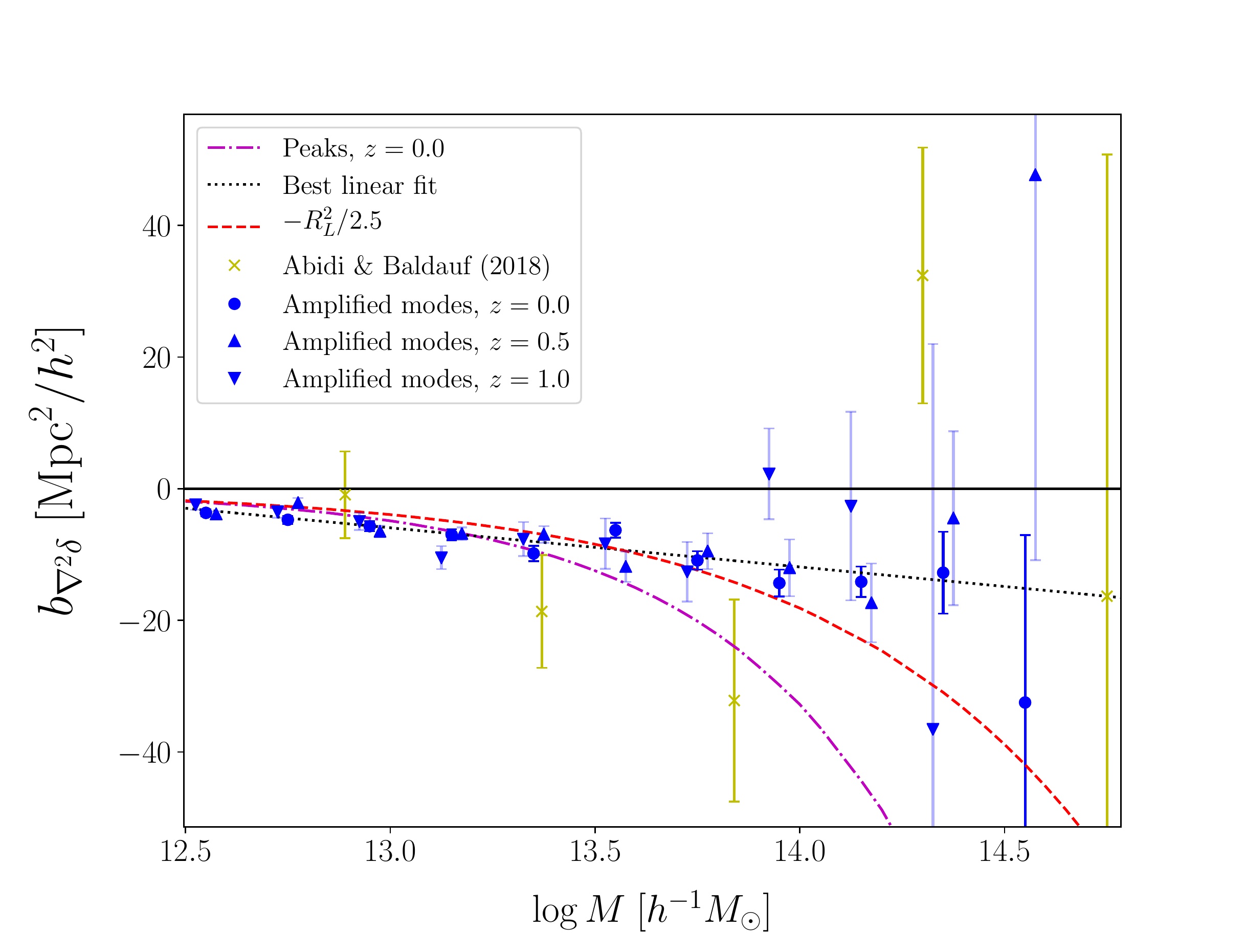}
\caption{$\blapl$ as a function of halo mass. The blue symbols show results obtained with our amplified-mode simulations (\refeq{blaplfinal}) at $z=0$, 0.5 and 1. The errorbars show the propagated $1\sigma$ bootstrap error of each term entering this equation. The yellow crosses are the measurements presented in \cite{Abidi:2018}. The dashed red and dotted-dashed magenta lines show $-R_L^2/2.5$ and the peak prediction respectively. Finally, we also provide a best fit linear in $\log M$ (black dotted line). See text for more discussion.}
\label{fig:blapl}
\end{figure}

\refFig{blapl} shows results for $b_{\nabla^2\d}$  as a function of halo mass at redshift 0, 0.5 and 1 (blue circles, upper triangles and lower triangles respectively). We obtain a clear detection of a nonzero, negative bias at all mass, with the most precise constraints being at $z=0$. Interestingly, no strong evidence of a redshift dependence is seen in the relation $\blapl(M)$. This is in strong contrast to other bias parameters, such as $b_1$, which are strongly redshift-dependent at fixed mass (they are approximately universal functions of $\sigma(M,z)$). 
For this reason, the fits described below are performed to the combined results at all redshifts. Notice that we do not have results for $\log M=14.55$ at $z=1$ since at this redshift the number of objects is too low to obtain a robust measurement.

We compare our results with the results from \citep{Abidi:2018} who obtained their measurement from a fit to the 1-loop halo-matter power spectrum (as described in \refsec{powerspec}) at $z=0$ only. We find consistent results within errors. The measurements reported here have smaller error bars and correspondingly higher signal to noise. We also obtained results from a fit to the 1-loop halo-matter power spectrum but with a very low signal-to-noise ratio. Hence we present them in \reffig{blaplcomp} in \refapp{Phm1loop} as a cross-check of our results. In addition we performed two fits of our combined results at all redshifts. Firstly, we fitted the relation $-R^2_L(M)/\alpha$ (the Lagrangian radius squared divided by some constant). Indeed, as explained in \refsec{intro}, since $\blapl$ is the first dimensionful bias parameter, we expect it to have a magnitude of the order of the only physical scale entering halo formation, i.e. $R^2_L$. We obtained a best fit value
\be
\blapl (M) \sim -\frac{R_L^2(M)}{2.5} \, .
\label{eq:RLfit}
\ee
This relation is shown by the red dashed curve in \reffig{blapl}. We see that this performs quite well, confirming the expected trend. However we obtained a reduced $\chi^2$ per degree of freedom $\chi^2=15$ which is quite poor; there are also deviations from a $R_L^2(M)$ scaling both at low and high masses. Hence we also  performed a simple linear fit in $\log M$ shown by the black dotted curved. We obtained
\be
\blapl (M) = -5.9 \log M + 71.3 \, (\Mpch)^2\, ,
\label{eq:linfit}
\ee
with a reduced $\chi^2=5.6$ hence performing much better (we checked that a quadratic fit does not improve the $\chi^2$). The physical interpretation of this logarithmic mass dependence however is unclear. Hence we caution against using this fitting relation outside the mass and redshift range probed here.

Finally, we compare our results with the peak prediction described in the previous section (magenta dashed-dotted line). This analytic prediction is the one that performs the worst quantitatively, especially at higher mass. However it is satisfying to see that this fairly simple model predicts the correct sign for this bias parameter, and a roughly correct overall amplitude. Moreover, it only predicts a weak redshift dependence of $\blapl$ at fixed mass, which is confirmed by our results.

%%%%%%%%%%%%%%%%%%%%%%%%%%%%%%%%%%%%%%%%%%%%%%%%%%%%%%%%%%%%%%%%%%%%%%%%%%%%%
%%%%%%%%%%%%%%%%%%%%%%%%%%%%%%%%%%%%%%%%%%%%%%%%%%%%%%%%%%%%%%%%%%%%%%%%%%%%%
\section{Conclusions}
\label{sec:concl} 

Amplified-mode simulations are a simple and efficient way to effectively modify the Laplacian of the matter density field $\dm$ on large scales by amplifying the real part of a Fourier mode at a given wavenumber $k_0$. This corresponds to adding a cosine wave in configuration space. We have performed various tests and computations in \refapp{checksims} and \refapp{deltaandP} to validate our implementation.

Using these simulations, and taking into account the nontrivial coupling between the amplified mode and short-wavelength modes, we have presented new measurements of the halo higher-derivative bias parameter $\blapl$. This allowed us to obtain a clear detection of this parameter for all halo masses probed and up to $z=1$. Our results, which are the most robust to date, are in broad agreement with those of \cite{Abidi:2018} with however a much higher signal-to-noise ratio.

We obtained a negative bias parameter at all mass with almost no redshift dependence in agreement with the analytic prediction from peak theory. Since $\blapl$ has dimension of [length]$^2$, it involves a physical length scale which, in addition to the nonlinear scale $k_{\rm NL}$, determines the range of scales on which rigorous perturbative approaches can be trusted. 
On physical grounds, we expect its magnitude to be of the order of the physical length scale in halo formation, the halo Lagrangian radius $R_L$. We indeed found $-R_L^2/2.5$ to be a reasonably good fit to our results. Our measurements thus confirm that nonlocal effects should only become important on scales $\lesssim R_L$. We also provided a simple empirical fit which performs better than the scaling with $R_L$ and that can be used quantitatively in future work in \refeq{linfit}. It should only be trusted over the mass and redshift range probed here, however.

These measurements complete the ones of \cite{Lazeyras:2015, Lazeyras:2017}, and we now have results for the complete set of bias parameters entering the 1-loop halo power spectrum and tree-level bispectrum for the cosmology adopted in this work at $z=0,$ 0.5 and 1. This enables future work such as studying the reach of perturbation theory by comparing, e.g. the halo-matter power spectrum as measured in simulation with the 1-loop SPT prediction with no free parameter. It will also allow us to study the stochasticity in halo formation in more detailed ways since we will be able to push the small scale-limit of the deterministic part (i.e. describe $\d_h$ and its associated statistics accurately to higher $k$ values). Finally, it could also be helpful for future surveys. Indeed, in order to be able to use the observed distribution of discrete luminous tracers on smaller and smaller scales (thus increasing constraining power) to extract cosmological information, it is of crucial importance to use a robust bias model with as few free parameters as possible. The measurements of this work together with those of \cite{Lazeyras:2015, Lazeyras:2017}, and the fitting functions provided, are precisely aiming towards this direction. 

\acknowledgments{We thank Tobias Baldauf for useful discussions at the early stages of this project, Vincent Desjacques for his insight regarding the peak model, and Mehrdad Mirbabayi for his help regarding higher-order corrections. TL would like to thank KITP and MPA for their hospitality during some of this work. This research was supported in part by the National Science Foundation under Grant No. NSF PHY-1125915. FS acknowledges support from the Starting Grant (ERC-2015-STG 678652) ``GrInflaGal'' from the European Research Council.
}

%%%%%%%%%%%%%%%%%%%%%%%%%%%%%%%%%%%%%%%%%%%%%%%%%%%%%%%%%%%%%%%%%%%%%%%%%%%%%
%%%%%%%%%%%%%%%%%%%%%%%%%%%%%%%%%%%%%%%%%%%%%%%%%%%%%%%%%%%%%%%%%%%%%%%%%%%%%
\appendix

\section{Simulation checks}
\label{app:checksims} 

This appendix presents a simple test to confirm that our implementation of the amplified-mode simulations is correct. We setup the initial condition of the simulation such that the final density field at redshift 0 is $\d(\vk, z=0)=\Delta \, (L^3/2) \,  \left[\d_D(\vk-\vk_0)+\d_D(\vk+\vk_0)\right]$, i.e. we set all modes of the density field to zero and only apply the mode amplification at $\vk=\vk_0$ (the factor $(L^3/2)$ comes from the fact that we want the amplitude to be $\Delta$ in real space and that we take discrete Fourier transforms, see \refapp{deltaandP}). We do this for a plane wave aligned with the $x$ axis, and various $\vk_0$, initial redshift, amplitude $\Delta$ and box size $L$. We then output the density field in real space in order to check that we recover a cosine wave at the right wavenumber and with the right amplitude. This also allows us to verify that the redshift dependence of $\Delta$ is indeed given by $D(z)$ (the linear growth factor). We also output the results at redshift 0 both when running the full N-body code or when simply applying the 2LPT algorithm (since we work on scales that are still linear today we expect the two to be equal). 

\begin{figure}
\centering
\includegraphics[scale=0.36]{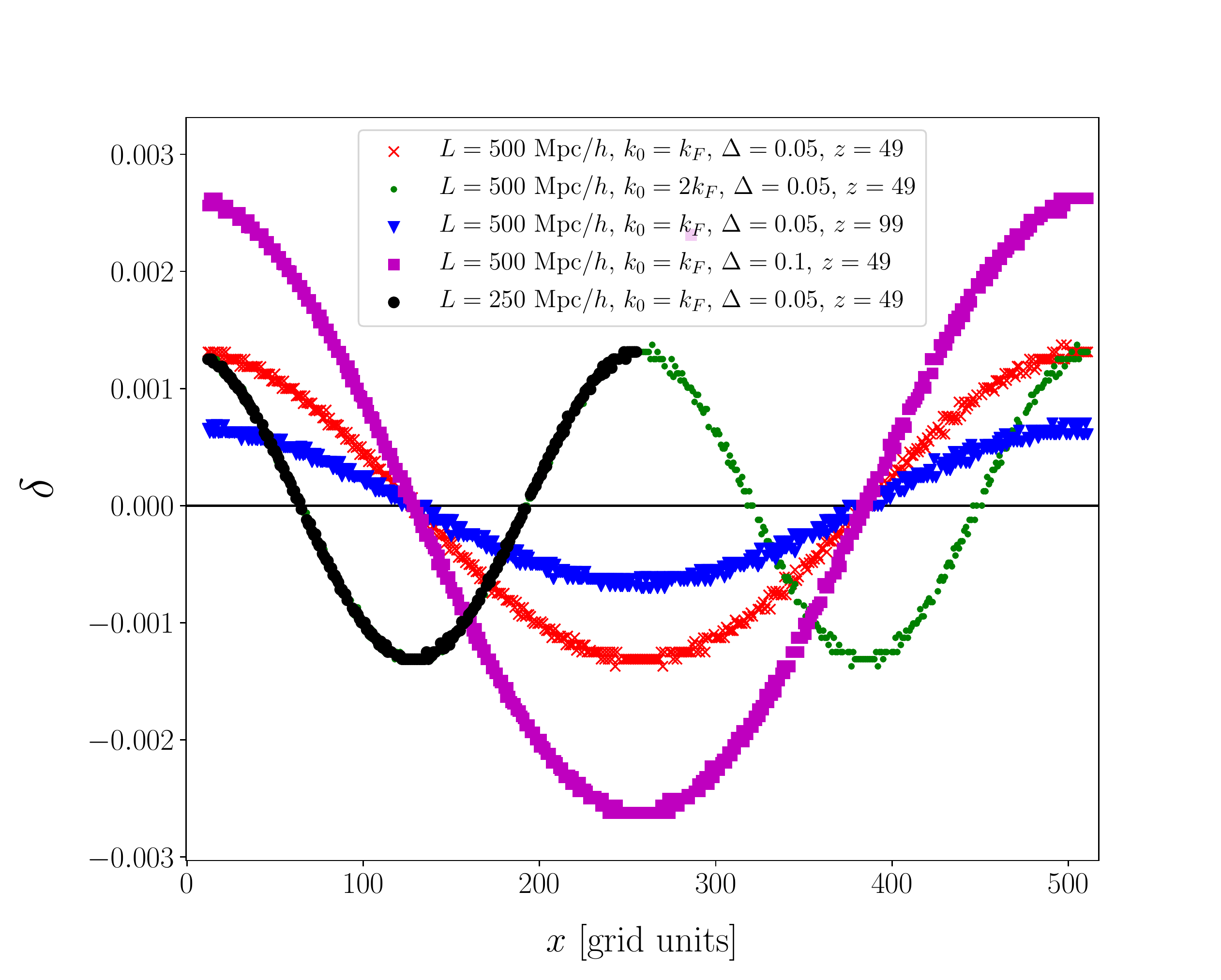}
\caption{Real space density as a function of $x$ (in grid units) in the initial conditions of an amplified-mode simulation with all modes set artificially to 0 and a single amplified mode (aligned with the $x$-axis as in the standard setup used in this work). The default setup is shown by the red crosses, i.e. redshift 49, box size $L=500 \Mpch$, linear amplitude at $z=0$ $\Delta=0.05$ and fundamental mode $k_F$ being amplified. We then show results varying each of these parameters. The green dots show $\d(x)$ when the mode $k_0=2k_F$ is amplified, the magenta squares are for an amplitude $\Delta=0.1$, blue triangles are at $z=99$, and the black circles are for a box size of $250 \Mpch$. See text for a discussion.}
\label{fig:deltaIC}
\end{figure}

Results are presented in \reffigs{deltaIC}{deltaz0}. In the first figure, we show the result of the 2LPT code used for the initial conditions at high redshift. The default setup is shown by the red crosses, i.e. output at redshift 49, box size $L=500 \Mpch$, linear amplitude at $z=0$ $\Delta=0.05$ and fundamental mode $k_F$ being amplified. We then show results varying each of these parameters. The green dots show $\d(x)$ when the mode $k_0=2k_F$ is amplified, the magenta squares are for an amplitude $\Delta=0.1$, blue triangles are at $z=99$, and the black circles are for a box size of $250 \Mpch$. 

\begin{figure}
\centering
\includegraphics[scale=0.45]{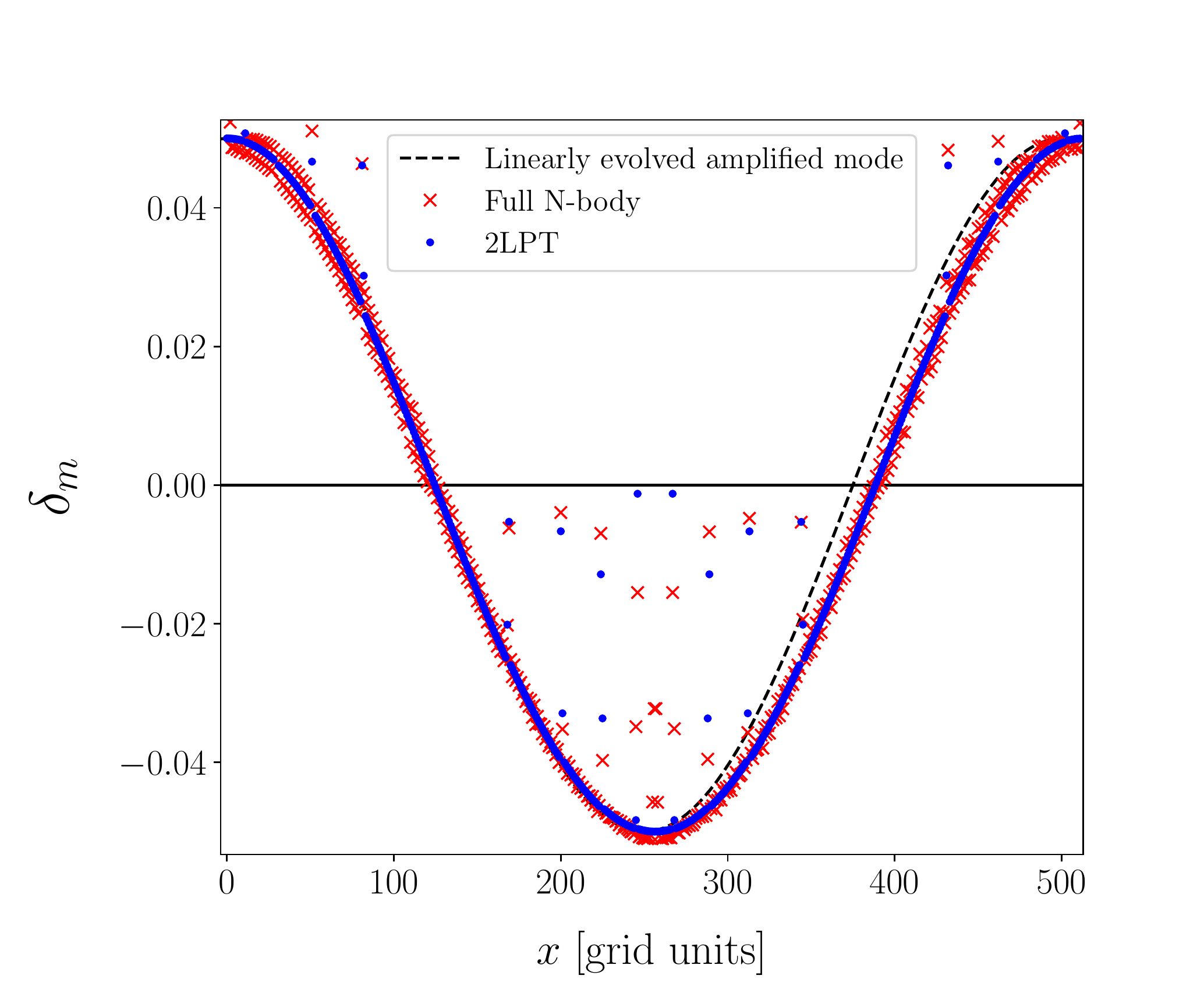}
\caption{Same as \reffig{deltaIC} but at redshift 0. The red crosses show the result when the full N-body code is run, while the result of the 2LPT code is shown as blue dots. We also show the expected result, i.e. the linearly evolved amplified mode as the black dashed line for comparison. See text for more discussion.}
\label{fig:deltaz0}
\end{figure}

The first point to note is that we do indeed retrieve a cosine wave at the required frequency, as shown by comparing the red and green curves. The amplitude of the wave is also correct since we obtain $\Delta(z=49)=0.0013$ and $\Delta(z=99)=0.00065$ when setting  $\Delta(z=0)=0.05$ (red and blue curves respectively). This is expected as we work on linear scales and hence the amplitude should evolve with the growth factor $D(z)$. For our cosmology $D=0.76, \, 0.02$ and $0.01$ at redshift 0, 49 and 99 respectively, and $D(z)/D(0) \cdot 0.05 = 0.0013 (0.00065)$ at $z=49 (99)$ respectively.  We furthermore get $\Delta(z=49)=0.0026$ when setting $\Delta(z=0)=0.1$ (magenta curve) which is indeed twice the amplitude of the red curve. Finally, when dividing the box size by two and amplifying the fundamental mode we get the black circles which is also the correct result since the fundamental mode of this box corresponds to $2k_F$ for $L=500 \Mpch$.

We now turn to \reffig{deltaz0}, which shows the same as the red curve in \reffig{deltaIC} but at redshift 0. The red crosses show the result when the full N-body code is run. We see that we recover a cosine wave with the expected amplitude at redshift zero, $\Delta=0.05$, which is indicated by the black dashed line showing the linearly evolved density field. A small number of grid cells display outlying values, which are also present when generating the density field at redshift zero using the 2LPT code (blue points). The outliers predominantly happen in the low-density region and are presumably due to inaccuracies in the force calculation in a few specific locations. In addition, the simulation results show small oscillations at a high wavenumber close to the Nyquist frequency of the particle grid. Since they do not appear in the 2LPT results, these must be due to the N-body algorithm. We also see the same effect when outputting the results from Gadget-2 at $z=48$, close to the initial redshift. Notice that in this test setup, there are no small-scale initial density perturbations. In a cosmological N-body simulation, which has significant initial density perturbations on small scales, we expect this numerical artifact to become irrelevant. 

\section{Density field in amplified-mode simulations at 1-loop order}
\label{app:deltaandP}
\label{app:1loopdelta} 

In this appendix we present the detailed computation of the matter and halo density fields at 1-loop in amplified-mode simulations using SPT. We omit the redshift argument throughout since these computations are valid at all redshift. 

We start with the matter density field $\dm$. This will be useful to compute the overdensity of halos later. We start again from the expression for the linear density field in the amplified-mode simulations \refeq{d1mod} with $\phi=0$ without loss of generality.
We then consider the nonlinear density field. In perturbation theory, it can be written as
\ba
\d_m(\vk) =\:& \left[1 - C^2_{s, \, \rm eff} k^2 \right] \d^{(1)}(\vk) \vs
& + \sum_{n=2}^\infty \int_{\vp_1} \cdots \int_{\vp_n} (2\pi)^3 \d_D(\vk - \vp_{1\cdots n})F_n(\vp_1,\cdots,\vp_n) \d^{(1)}(\vp_1)\cdots \d^{(1)}(\vp_n)\,,
\label{eq:dPTm}
\ea
where $F_n$ are the fully symmetrized kernels of the matter density field. We now insert \refeq{d1mod} into \refeq{dPTm} and evaluate the result up to cubic order. This gives
\ba
\d_m(\vk) \:& = \left[1 - C^2_{s, \, \rm eff} k^2 \right] \left(\d_s^{(1)}(\vk)+\frac12 \Delta (2\pi)^3\left[ \d_D(\vk-\vk_0) + \d_D(\vk+\vk_0)\right]\right) \vs
& + \int_{\vp_1} \int_{\vp_2} (2\pi)^3 \d_D(\vk - \vp_{1 2})F_2(\vp_1,\vp_2) \d_s^{(1)}(\vp_1)\d_s^{(1)}(\vp_2) \vs
& + 2\int_{\vp_1} \int_{\vp_2} (2\pi)^3 \d_D(\vk - \vp_{1 2})F_2(\vp_1,\vp_2) \d_s^{(1)}(\vp_1)\frac12 \Delta (2\pi)^3\left[ \d_D(\vp_2-\vk_0) + \d_D(\vp_2+\vk_0)\right] \vs
& + \int_{\vp_1} \int_{\vp_2}(2\pi)^3 \d_D(\vk - \vp_{12})F_2(\vp_1,\vp_2) \frac14 \Delta^2 (2\pi)^6 \left[ \d_D(\vp_1-\vk_0) + \d_D(\vp_1+\vk_0)\right]\vs
& \hspace{2.2cm} \times \left[ \d_D(\vp_2-\vk_0) + \d_D(\vp_2+\vk_0)\right]\vs
& + \int_{\vp_1} \int_{\vp_2} \int_{\vp_3} (2\pi)^3 \d_D(\vk - \vp_{1 23})F_3(\vp_1,\vp_2,\vp_3) \d_s^{(1)}(\vp_1)\d_s^{(1)}(\vp_2) \d_s^{(1)}(\vp_3)\vs
& + 3 \int_{\vp_1} \int_{\vp_2} \int_{\vp_3} (2\pi)^3 \d_D(\vk - \vp_{1 23})F_3(\vp_1,\vp_2,\vp_3) \d_s^{(1)}(\vp_1)\d_s^{(1)}(\vp_2) \vs
& \hspace{2.5cm} \times \frac12 \Delta (2\pi)^3\left[ \d_D(\vp_3-\vk_0) + \d_D(\vp_3+\vk_0)\right] \vs
& + 3 \int_{\vp_1} \int_{\vp_2} \int_{\vp_3} (2\pi)^3 \d_D(\vk - \vp_{1 23})F_3(\vp_1,\vp_2,\vp_3) \d_s^{(1)}(\vp_1)\frac14 \Delta^2 (2\pi)^6 \vs
& \hspace{2.5cm} \times \left[ \d_D(\vp_2-\vk_0) + \d_D(\vp_2+\vk_0)\right]\left[ \d_D(\vp_3-\vk_0) + \d_D(\vp_3+\vk_0)\right] \vs
& + \int_{\vp_1} \int_{\vp_2} \int_{\vp_3} (2\pi)^3 \d_D(\vk - \vp_{1 23})F_3(\vp_1,\vp_2,\vp_3) \frac18 \Delta^3 (2\pi)^9 \left[ \d_D(\vp_1-\vk_0) + \d_D(\vp_1+\vk_0)\right]\vs
& \hspace{2.2cm} \times \left[ \d_D(\vp_2-\vk_0) + \d_D(\vp_2+\vk_0)\right]\left[ \d_D(\vp_3-\vk_0) + \d_D(\vp_3+\vk_0)\right]\,,
\label{eq:dPTcubic}
\ea
where we have used the symmetry of the kernels under permutation of their argument. We are interested in the response of the nonlinear matter field to the long-wavelength enhanced mode for which the estimator is defined as the symmetric difference
\be
\widehat{\frac{{\rm d} \d_m}{{\rm d}\Delta}} (\vk) \equiv \frac{1}{2\Delta} \left[
\d_m(\vk)\Big|_{+\Delta} - \d_m(\vk)\Big|_{-\Delta}
\right]\,.
\label{eq:estimatormatter}
\ee
By symmetry, only terms of \refeq{dPTcubic} that are odd in $\Delta$ contribute to this. Since we typically choose $\Delta \sim 10^{-2} - 10^{-1}$ we further neglect the terms that are cubic in $\Delta$ and we obtain
\ba
\widehat{\frac{{\rm d} \d_m}{{\rm d}\Delta}}(\vk) \:& = \left[1 - C_{s,\rm eff}^2 k^2 + \O(k^4) \right] \frac12 \, (2\pi)^3\left[\d_D(\vk-\vk_0)+\d_D(\vk+\vk_0)\right] \vs
& + \left[F_2(\vk-\vk_0 ,\vk_0)\d^{(1)}_s(\vk-\vk_0)+F_2(\vk+\vk_0 ,-\vk_0)\d^{(1)}_s(\vk+\vk_0)\right] \vs
&+ \frac32 \int_{\vp}F_3(\vp,\vk-\vk_0-\vp,\vk_0) \d_s^{(1)}(\vp)\d_s^{(1)}(\vk-\vk_0-\vp) \vs
& + \frac32 \int_{\vp}F_3(\vp,\vk+\vk_0-\vp,-\vk_0) \d_s^{(1)}(\vp)\d_s^{(1)}(\vk+\vk_0-\vp)\,.
\ea
We now average over many realizations. Terms with odd powers of $\d_s^{(1)}$ average to zero, and are hence dropped as well, while terms with even powers of $\d_s^{(1)}$ can be replaced by their ensemble average
\be
\< \d_s^{(1)}(\vp)\d_s^{(1)}(\vp') \> = (2\pi)^3 \d_D(\vp+\vp') \Plin(p)\,.
\label{eq:dlinav}
\ee
In the end, only terms that involve zero power of $\Delta$ and two powers of $\d_s^{(1)}$
remain. For the first, there are terms in which the arguments of $\d_s^{(1)}$ sum to zero, and those where they sum to $2\vk_0$ when evaluating the expression at $\vk_0$. The latter terms average to zero when considering many realizations since $\d_s^{(1)}$ obeys  \refeq{dlinav} by definition, and we hence get
\ba
\left \langle \widehat{\frac{{\rm d} \d_m}{{\rm d}\Delta}} \right \rangle (\vk) =\:& \left[1 - C_{s,\rm eff}^2 k^2 + \O(k^4) \right] \frac12 \, (2\pi)^3\left[\d_D(\vk-\vk_0)+\d_D(\vk+\vk_0)\right] \vs
& + \frac32 \int_{\vp} (2\pi)^3 \d_D(\vk-\vk_0)\Plin(p)F_3(\vp,\vk-\vk_0-\vp,\vk_0)+ \O(\Delta^2) \,.
\label{eq:dm3d}
\ea
Finally, evaluating this expression at $\vk=\vk_0$ leads to
\ba
\left \langle \widehat{\frac{{\rm d} \d_m}{{\rm d}\Delta}} \right \rangle(\vk_0) =\:& \left[1 -  C_{s,\rm eff}^2 k_0^2 + \O(k_0^4) \right] \frac12 \, (2\pi)^3\d_D(\v{0}) \vs
& + \frac32 \int_{\vp} \Plin(p) F_3(\vp,-\vp,\vk_0) (2\pi)^3\d_D(\v{0}) + \O(\Delta^2) \,.
\label{eq:dm3dPS}
\ea
Now everything is multiplied by the same factor $(2\pi)^3 \d_D(\v{0})$ (which simply gives  $L_{\rm box}^3$ when restoring box normalization).

\begin{figure}
\centering
\includegraphics[scale=0.45]{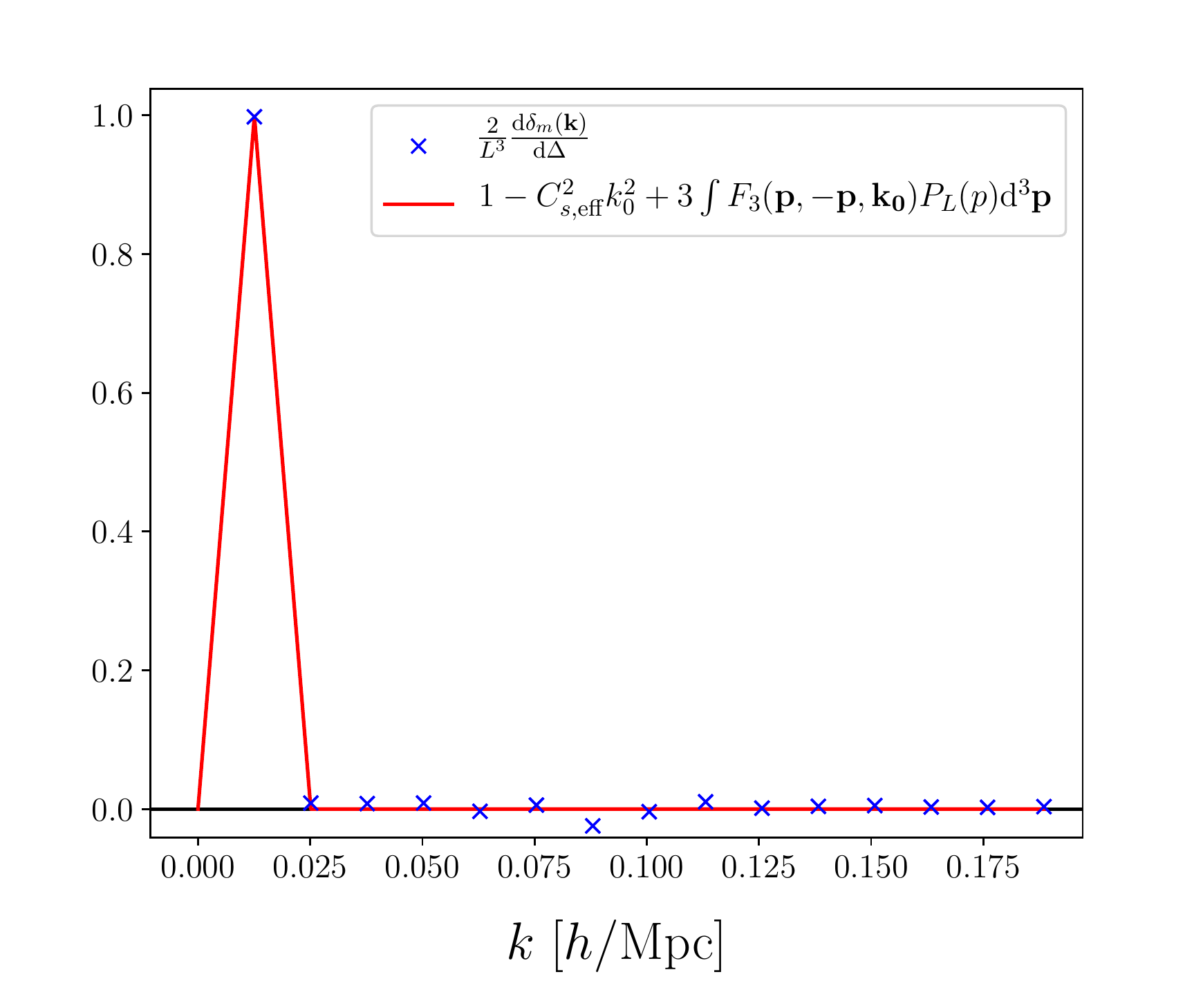}
\caption{Comparison between the measured response of the matter density field to an amplification of the fundamental mode of the box averaged over all realizations, and the SPT prediction as a function of $k$ at z=0.  The blue crosses are the l.h.s of \refeq{dm3d} measured in the amplified-mode simulations, while the red line is the r.h.s. As expected, only the mode $\dm(k_0)$ responds to the amplification after averaging over realizations.}
\label{fig:ddmodDofk}
\end{figure}

This last expression implies that it would be possible to measure $\cseff$ from amplified-mode simulations by looking at the matter field response to the mode amplification. For simplicity, and in order to provide a more direct comparison with the literature on $\cseff$, we measure this coefficient from the matter power spectrum. We however verified that \refeq{dm3d} is respected in our simulations. This is shown in \reffig{ddmodDofk} at redshift 0 and for $k_0=k_F$. The blue crosses represent the l.h.s of \refeq{dm3d} while the red line is the right one. We clearly see that the Dirac deltas on the r.h.s of the equation prevent modes at $k \neq k_0$ to respond to the amplification. More quantitatively, the value of the blue cross at $k=k_F$ is $0.9979 \pm 5\cdot 10^{-4}$ while the red line is $0.9986 \pm 2 \cdot 10^{-4}$.

We now move to the halo density field and present details of the computation leading to \refeq{blaplfinal} not shown in the main text. The computation follows the same line as for the matter case. We start from \refeq{estimator} and insert into it the nonlinear halo density field up to third order as given by \refeq{dPT}. Following the same reasoning as to go from \refeq{estimatormatter} to \refeq{dm3d} yields the terms proportional to $\Delta$ (before dividing by $\Delta$ in the estimator). For the terms proportional to $\Delta^3$, we have
\ba
\frac18 \Delta^3 \sum_{r,s,t = -1, 1} (2\pi)^3 \, \frac12 & \Big[F_3(r\vk_0,s\vk_0,t\vk_0)\d_D(\vk_0 - (r+s+t)\vk_0) \vs 
& + \d_D(-\vk_0 - (r+s+t)\vk_0 ) F_3(r\vk_0,s\vk_0,t\vk_0)\Big]\,.
\ea
The factor $\Delta^3/8$ comes from the last two terms in \refeq{d1mod}, while the square brackets contain the two terms in \refeq{estimator}, taking into account that $(-\Delta)^3 = -\Delta^3$. It is easy to see that only 3 combinations of $r,s,t$ contribute to teach term, so we obtain
\be
\widehat{\frac{{\rm d} \d_h}{{\rm d}\Delta}}\Big|_{\Delta^3} = \frac38 (2\pi)^3 \d_D(\v{0})\Delta^2 F^h_3(\vk_0,\vk_0,-\vk_0)\,,
\ee
using the symmetries of the fully symmetrized $F_3$ kernel. Hence we get
\ba
\widehat{\frac{{\rm d} \d_h}{{\rm d}\Delta}} =\:& \left[b_1 - (b_{\lapl\d} + b_1 C_{s,\rm eff}^2) k_0^2 + \O(k_0^4) \right] \frac12 \, (2\pi)^3\d_D(\v{0}) \vs
& + \frac32 \int_{\vp} \d^{(1)}_s(\vp)\d^{(1)}_s(-\vp)F_3^{(h)}(\vp,-\vp,\vk_0) \vs
& + \frac38 (2\pi)^3 \d_D(\v{0})\Delta^2 F_3^{(h)}(\vk_0,-\vk_0,\vk_0) \,.
\label{eq:dh3d}
\ea
In the last line we have used the symmetry of the perturbation theory kernels under sign change of all momenta $F_3^{(h)}(-\vk_1,-\vk_2,-\vk_3) = F_3^{(h)}(\vk_1,\vk_2,\vk_3)$. Note that there are no odd contributions in $\Delta$ at second order, and that $\d^{(1)}_s(\vp)\d^{(1)}_s(-\vp) = |\d_s^{(1)}(\vp)|^2$ is positive definite. When averaging over many small scale modes we can replace 
 \be
\d^{(1)}_s(\vp)\d^{(1)}_s(-\vp) \to \<\d^{(1)}_s(\vp)\d^{(1)}_s(-\vp)\> = (2\pi)^3 \d_D(\v{0}) \Plin(p)\,, 
\ee
which finally leads to \refeq{dh3dPS}.

\section{On the effective sound speed of matter}
\label{app:cseff}

\begin{figure}
\centering
\includegraphics[scale=0.5]{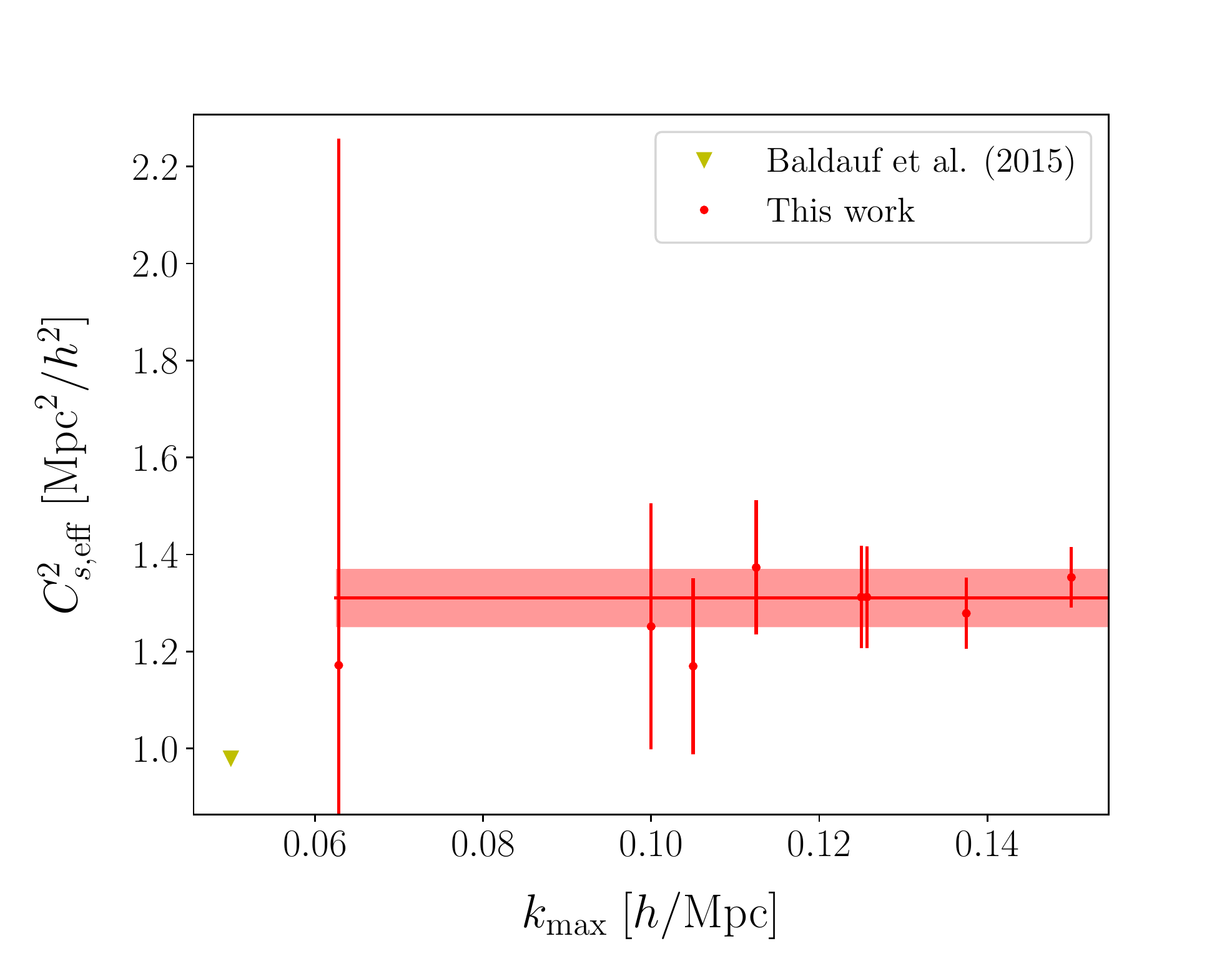}
\caption{$\cseff$ as a function of $k_{\rm max}$ used for the fit in \refeq{cPSPT} at redshift 0 (red points).  The red solid line shows the mean value inferred from results in the range $k_{\rm max} \in [0.065,0.15] \, \iMpch$. The shaded region shows the 1$\sigma$ error. We compare it to results from EFT from \cite{Baldauf:2015aha} (yellow triangle). We find no significant dependence of $\cseff$ with the maximum $k$ used for the fit up to $k_{\rm max}=0.15 \iMpch$. At higher $k_{\rm max}$ we found the same observed scale dependence as was also found by \citep{Baldauf:2015aha} who showed that it is due to 2-loop terms that are not taken into account in our estimator \refeq{cPSPT}, which is also why we restrict ourselves to $k \leq 0.15 \iMpch$. See text for more details.}
\label{fig:cseffcomp}
\end{figure}

This appendix presents further results and tests on $\cseff$. In \reffig{cseffcomp} we show the results for $\cseff$ as a function of $k_{\rm max}$ used for the fit in \refeq{cPSPT}. We also show the mean value inferred from results with $k_{\rm max}$ in the range $[0.063,0.15] \, \iMpch$.  We decided to use the mean value in  this range since at higher values of $k_{\rm max}$ the central value shifts systematically which is most likely due to higher-order effects (2-loop terms) that we do not take into account. As we discuss in the next paragraph, \cite{Baldauf:2015aha} already found these effects to be important. We hence obtain $\cseff=1.31 \pm 0.06 \, (\Mpch)^2$ at $z=0$. 

We compared this to results the results from \cite{Carrasco:2013sva}, \cite{ Angulo:2015}, and \cite{Baldauf:2015aha}. These authors respectively quote values of $1.6$, $2.31\pm 0.02$, and $0.98 \, (\Mpch)^2$ at $z=0$. There are thus discrepancies between published results in the literature.
In the case of \cite{Carrasco:2013sva} and \cite{Angulo:2015}, this could be due to the fact that they used a similar approach than us but with a single fit in the range $0.15-0.25 (0.3) \iMpch$ respectively, where 2-loop terms induce a scale dependence in $\cseff$. The most interesting result was found by \citep{Baldauf:2015aha} who found a result close to our one using the same technique as in this work with $k_{\rm max}=0.05 \iMpch$. Their figure 6 is similar to our \reffig{cseffcomp} at various redshifts and up to higher $k$. We found a very similar scale dependence of $\cseff$ than them when we looked at similar $k_{\rm max}$. These authors showed that this is due to 2-loop terms that are not taken into account in our estimator \refeq{cPSPT}, and they attribute to this the discrepancy with other works. On the other hand, as can be seen on \reffig{cseffcomp}, we do not see any strong dependence of $\cseff$ with the maximum $k$ used for the fit up to $k_{\rm max}=0.15 \iMpch$ which is why we limit ourselves to this range. Notice that this is a bit larger than, but comparable to the maximum wavenumber used to measure $\blapl$, which is $0.126\iMpch$ (or less, depending on mass). 
  Some disagreement with values reported in the literature is also expected due to the fact that we use different cosmologies. However, the change in $\cseff$ due to a change a cosmology should be mainly proportional to the amplitude of the linear power spectrum, i.e. $\sigma_8^2$, leading to expected differences only of order 10\%. 

\begin{figure}
\centering
\includegraphics[scale=0.5]{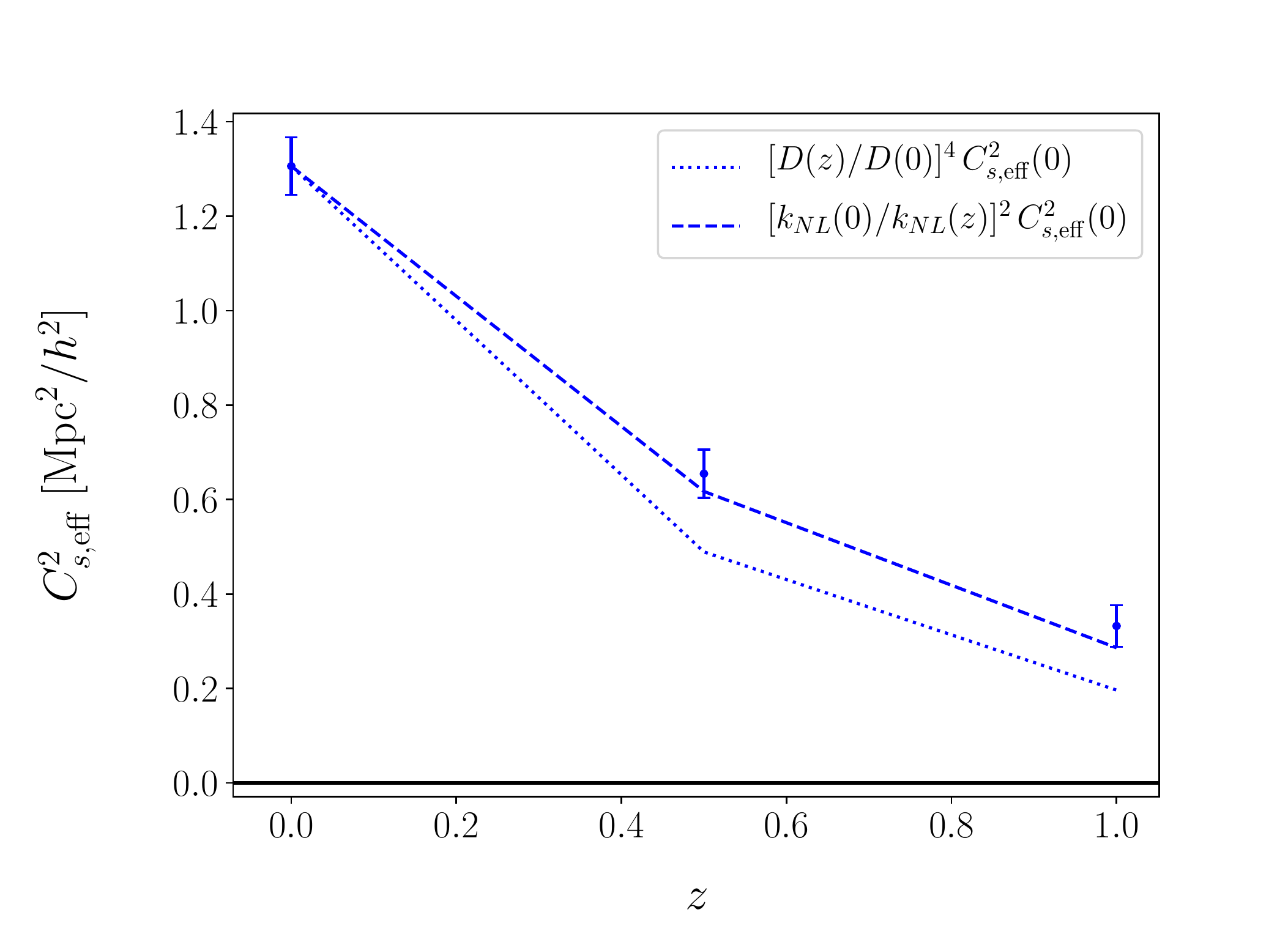}
\caption{Redshift dependence of the effective sound speed of matter  $C_{s,{\rm eff}}^2$. The blue points show results from the 1-loop power spectrum fit at $z=0, \, 0.5,$ and $1$ with 1$\sigma$ errorbars. The dotted line shows a $D^4$ dependence while the dashed one shows the behavior of $1/k_{\rm NL}^2$, which are the expected results (see right panel of figure 11 in \citep{Baldauf:2015tla}).}
\label{fig:cofkzdep}
\end{figure}

Finally, we also look at the redshift dependence of $\cseff$ in \reffig{cofkzdep}. In order to cancel divergences in perturbation-theory loop integrals, $\cseff$ has to scale as $D^4(z)$, although the finite part which remains after the divergences are canceled and which is the parameter we are measuring could scale differently with redshift. In scale-free cosmologies (flat, matter-dominated cosmology with a power-law power spectrum), one expects the finite part to scale as $k_{\rm NL}^{-2}$ following dimensional reasoning.  Indeed the $1/k_{\rm NL}^2$ relation reproduces the redshift dependence we find well, while the scaling $D^4$ is a bit too steep.

\section{Comparison with results from the 1-loop halo-matter power spectrum}
\label{app:Phm1loop} 

\begin{figure}
\centering
\includegraphics[scale=0.55]{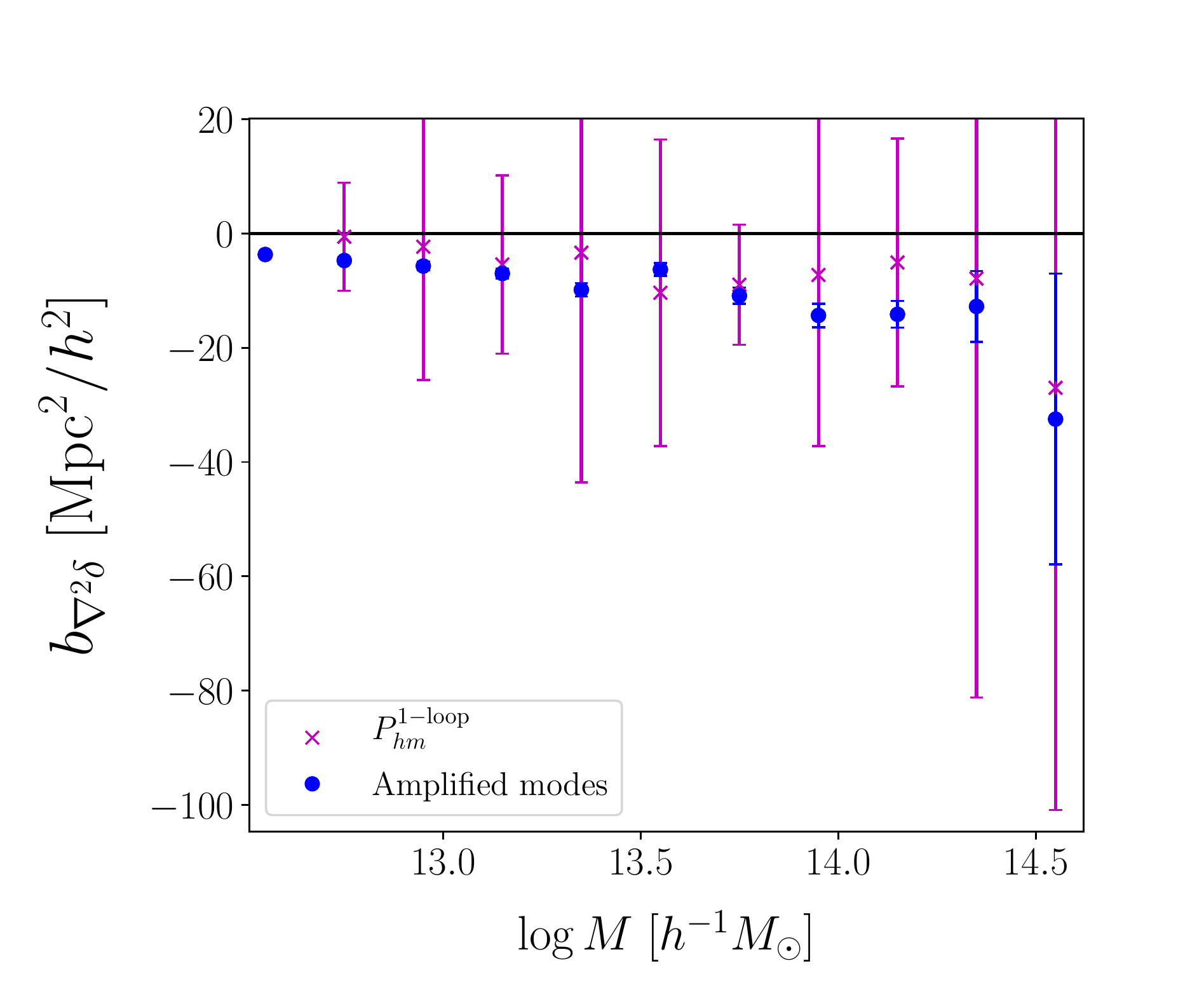}
\caption{$\blapl$ as a function of halo mass $M$ at $z=0$. The blue symbols present results from amplified-mode simulations while the purple ones are from the 1-loop power spectrum.}
\label{fig:blaplcomp}
\end{figure}

\refFig{blaplcomp} presents a comparison between the measurements presented in \refsec{results} obtained from amplified-mode simulations, and measurements obtained from a fit to the 1-loop halo-matter power spectrum with one free parameter, as presented in \refsec{powerspec}. The results are at redshift 0. We see that we find a very good agreement between the two. However the signal-to-noise ratio is much higher for the results obtained from amplified-mode simulations which allow for a clear detection of $\blapl$. Notice that we performed the fit of the power spectrum up to $k_{\rm max}=0.15 \iMpch$, which roughly corresponds to $12 k_F$, and we use the L2400 set of simulations. Errorbars are the 1$\sigma$ error obtained following the procedure outlined in \refsec{powerspec}. We also emphasize that we use the CAMB linear power spectrum to compute the integrals multiplying $b_2$, $b_{K^2}$ and $\left(b_{K^2}+ 2/5 \, b_{\rm td}\right)$, and use these as mean values (as well as the mean values presented in \citep{Lazeyras:2017} for the bias parameters) when fitting \refeq{Phm1loop}. Cosmic variance is partially canceled by measuring $P_{\rm L}$ from the Zel'dovich density field at $z=99$ for each realization of L500, and using it in the first line of \refeq{Phm1loop} (also in the computation of $P_{mm}^{1-{\rm loop}}$). Using $P_{\rm L}$ and the bias values as measured in each realization for each term in \refeq{Phm1loop}, and performing a fit realization by realization would allow to cancel more cosmic variance.

\clearpage

%%%%%%%%%%%%%%%%%%%%%%%%%%%%%%%%%%%%%%%%%%%%%%%%%%%%%%%%%%%%%%%%%%%%%%%%%%%%
%%%%%%%%%%%%%%%%%%%%%%%%%%%%%%%%%%%%%%%%%%%%%%%%%%%%%%%%%%%%%%%%%%%%%%%%%%%%

\bibliography{references}

\providecommand{\href}[2]{#2}\begingroup\raggedright\begin{thebibliography}{10}

\bibitem{biasreview}
V.~Desjacques, D.~Jeong, and F.~Schmidt, {\it {Large-Scale Galaxy Bias}},
  \href{http://arxiv.org/abs/1611.09787}{{\tt arXiv:1611.09787}}.

\bibitem{Matsubara:1999}
T.~Matsubara, {\it {Stochasticity of bias and nonlocality of galaxy formation:
  Linear scales}},  {\em Astrophys. J.} {\bf 525} (1999) 543--553,
  [\href{http://arxiv.org/abs/astro-ph/9906029}{{\tt astro-ph/9906029}}].

\bibitem{Coles:2007}
P.~Coles and P.~Erdogdu, {\it {Scale-dependent Galaxy Bias}},  {\em JCAP} {\bf
  0710} (2007) 007, [\href{http://arxiv.org/abs/0706.0412}{{\tt
  arXiv:0706.0412}}].

\bibitem{Bardeen:1985}
J.~M. Bardeen, J.~R. Bond, N.~Kaiser, and A.~S. Szalay, {\it {The Statistics of
  Peaks of Gaussian Random Fields}},  {\em Astrophys. J.} {\bf 304} (1986)
  15--61.

\bibitem{Scherrer:1997}
R.~J. Scherrer and D.~H. Weinberg, {\it {Constraints on the effects of
  locally-biased galaxy formation}},  {\em Astrophys. J.} {\bf 504} (1998)
  607--611, [\href{http://arxiv.org/abs/astro-ph/9712192}{{\tt
  astro-ph/9712192}}].

\bibitem{Mann:1997}
B.~Mann, J.~Peacock, and A.~Heavens, {\it {Eulerian bias and the galaxy density
  field}},  {\em Mon. Not. Roy. Astron. Soc.} {\bf 293} (1998) 209--221,
  [\href{http://arxiv.org/abs/astro-ph/9708031}{{\tt astro-ph/9708031}}].

\bibitem{Fujita:2016}
T.~Fujita, V.~Mauerhofer, L.~Senatore, Z.~Vlah, and R.~Angulo, {\it {Very
  Massive Tracers and Higher Derivative Biases}},
  \href{http://arxiv.org/abs/1609.00717}{{\tt arXiv:1609.00717}}.

\bibitem{Angulo:2015}
R.~Angulo, M.~Fasiello, L.~Senatore, and Z.~Vlah, {\it {On the Statistics of
  Biased Tracers in the Effective Field Theory of Large Scale Structures}},
  {\em JCAP} {\bf 1509} (2015), no.~09 029,
  [\href{http://arxiv.org/abs/1503.08826}{{\tt arXiv:1503.08826}}].

\bibitem{Abidi:2018}
M.~M. Abidi and T.~Baldauf, {\it {Cubic Halo Bias in Eulerian and Lagrangian
  Space}},  {\em JCAP} {\bf 1807} (2018), no.~07 029,
  [\href{http://arxiv.org/abs/1802.07622}{{\tt arXiv:1802.07622}}].

\bibitem{Elia:2012}
A.~{Elia}, A.~D. {Ludlow}, and C.~{Porciani}, {\it {The spatial and velocity
  bias of linear density peaks and protohaloes in the {$\Lambda$} cold dark
  matter cosmology}},  {\em \mnras} {\bf 421} (Apr., 2012) 3472--3480,
  [\href{http://arxiv.org/abs/1111.4211}{{\tt arXiv:1111.4211}}].

\bibitem{Baldauf:2014}
T.~Baldauf, V.~Desjacques, and U.~Seljak, {\it {Velocity bias in the
  distribution of dark matter halos}},  {\em Phys. Rev.} {\bf D92} (2015)
  123507, [\href{http://arxiv.org/abs/1405.5885}{{\tt arXiv:1405.5885}}].

\bibitem{Paranjape:2013}
A.~Paranjape, E.~Sefusatti, K.~C. Chan, V.~Desjacques, P.~Monaco, and R.~K.
  Sheth, {\it {Bias deconstructed: Unravelling the scale dependence of halo
  bias using real space measurements}},  {\em Mon. Not. Roy. Astron. Soc.} {\bf
  436} (2013) 449--459, [\href{http://arxiv.org/abs/1305.5830}{{\tt
  arXiv:1305.5830}}].

\bibitem{Biagetti:2013}
M.~Biagetti, K.~C. Chan, V.~Desjacques, and A.~Paranjape, {\it {Measuring
  non-local Lagrangian peak bias}},  {\em Mon. Not. Roy. Astron. Soc.} {\bf
  441} (2014), no.~2 1457--1467, [\href{http://arxiv.org/abs/1310.1401}{{\tt
  arXiv:1310.1401}}].

\bibitem{Wagner:2014}
C.~Wagner, F.~Schmidt, C.-T. Chiang, and E.~Komatsu, {\it {Separate Universe
  Simulations}},  {\em Mon.Not.Roy.Astron.Soc.} {\bf 448} (2015) 11,
  [\href{http://arxiv.org/abs/1409.6294}{{\tt arXiv:1409.6294}}].

\bibitem{Lazeyras:2015}
T.~Lazeyras, C.~Wagner, T.~Baldauf, and F.~Schmidt, {\it {Precision measurement
  of the local bias of dark matter halos}},  {\em JCAP} {\bf 1602} (2016),
  no.~02 018, [\href{http://arxiv.org/abs/1511.01096}{{\tt arXiv:1511.01096}}].

\bibitem{assassi/etal}
V.~{Assassi}, D.~{Baumann}, D.~{Green}, and M.~{Zaldarriaga}, {\it
  {Renormalized halo bias}},  {\em \jcap} {\bf 8} (Aug., 2014) 56,
  [\href{http://arxiv.org/abs/1402.5916}{{\tt arXiv:1402.5916}}].

\bibitem{Lazeyras:2017}
T.~Lazeyras and F.~Schmidt, {\it {Beyond LIMD bias: a measurement of the
  complete set of third-order halo bias parameters}},  {\em JCAP} {\bf 1809}
  (2018), no.~09 008, [\href{http://arxiv.org/abs/1712.07531}{{\tt
  arXiv:1712.07531}}].

\bibitem{Baldauf:2015aha}
T.~Baldauf, L.~Mercolli, and M.~Zaldarriaga, {\it {Effective field theory of
  large scale structure at two loops: The apparent scale dependence of the
  speed of sound}},  {\em Phys. Rev.} {\bf D92} (2015), no.~12 123007,
  [\href{http://arxiv.org/abs/1507.02256}{{\tt arXiv:1507.02256}}].

\bibitem{abolhasani/mirbabayi/pajer:2016}
A.~{Akbar Abolhasani}, M.~{Mirbabayi}, and E.~{Pajer}, {\it {Systematic
  renormalization of the effective theory of Large Scale Structure}},  {\em
  \jcap} {\bf 5} (May, 2016) 063, [\href{http://arxiv.org/abs/1509.07886}{{\tt
  arXiv:1509.07886}}].

\bibitem{Gill:2004}
S.~P. Gill, A.~Knebe, and B.~K. Gibson, {\it {The Evolution substructure 1: A
  New identification method}},  {\em Mon.Not.Roy.Astron.Soc.} {\bf 351} (2004)
  399, [\href{http://arxiv.org/abs/astro-ph/0404258}{{\tt astro-ph/0404258}}].

\bibitem{Knollmann:2009}
S.~R. Knollmann and A.~Knebe, {\it {Ahf: Amiga's Halo Finder}},  {\em
  Astrophys.J.Suppl.} {\bf 182} (2009) 608--624,
  [\href{http://arxiv.org/abs/0904.3662}{{\tt arXiv:0904.3662}}].

\bibitem{Desjacques:2008jj}
V.~Desjacques, {\it {Baryon acoustic signature in the clustering of density
  maxima}},  {\em Phys. Rev.} {\bf D78} (2008) 103503,
  [\href{http://arxiv.org/abs/0806.0007}{{\tt arXiv:0806.0007}}].

\bibitem{Desjacques:2010:D81}
V.~{Desjacques} and R.~K. {Sheth}, {\it {Redshift space correlations and
  scale-dependent stochastic biasing of density peaks}},  {\em \prd} {\bf 81}
  (Jan., 2010) 023526, [\href{http://arxiv.org/abs/0909.4544}{{\tt
  arXiv:0909.4544}}].

\bibitem{Desjacques:2010}
V.~{Desjacques}, M.~{Crocce}, R.~{Scoccimarro}, and R.~K. {Sheth}, {\it
  {Modeling scale-dependent bias on the baryonic acoustic scale with the
  statistics of peaks of Gaussian random fields}},  {\em \prd} {\bf 82} (Nov.,
  2010) 103529, [\href{http://arxiv.org/abs/1009.3449}{{\tt arXiv:1009.3449}}].

\bibitem{Baldauf:2016}
T.~Baldauf and V.~Desjacques, {\it {Phenomenology of baryon acoustic
  oscillation evolution from Lagrangian to Eulerian space}},  {\em Phys. Rev.}
  {\bf D95} (2017), no.~4 043535, [\href{http://arxiv.org/abs/1612.04521}{{\tt
  arXiv:1612.04521}}].

\bibitem{Chan:2015}
K.~C. Chan, R.~K. Sheth, and R.~Scoccimarro, {\it {Effective Window Function
  for Lagrangian Halos}},  \href{http://arxiv.org/abs/1511.01909}{{\tt
  arXiv:1511.01909}}.

\bibitem{Carrasco:2013sva}
J.~J.~M. Carrasco, S.~Foreman, D.~Green, and L.~Senatore, {\it {The 2-loop
  matter power spectrum and the IR-safe integrand}},  {\em JCAP} {\bf 1407}
  (2014) 056, [\href{http://arxiv.org/abs/1304.4946}{{\tt arXiv:1304.4946}}].

\bibitem{Baldauf:2015tla}
T.~Baldauf, E.~Schaan, and M.~Zaldarriaga, {\it {On the reach of perturbative
  descriptions for dark matter displacement fields}},  {\em JCAP} {\bf 1603}
  (2016), no.~03 017, [\href{http://arxiv.org/abs/1505.07098}{{\tt
  arXiv:1505.07098}}].

\end{thebibliography}\endgroup
\end{document}